\documentclass[12pt]{article}
\pdfoutput=1

\usepackage{draft} 
\usepackage{hyperref}
\usepackage{graphicx,color,subfig}
\usepackage{cite}
\usepackage{mciteplus}
\usepackage{skak}
\usepackage{empheq}
\usepackage{anyfontsize}
\usepackage[normalem]{ulem}

\DeclareFontFamily{OT1}{pzc}{}
\DeclareFontShape{OT1}{pzc}{m}{it}{<-> s * [1.10] pzcmi7t}{}
\DeclareMathAlphabet{\mathpzc}{OT1}{pzc}{m}{it}

\def\be#1\ee{\begin{align}#1\end{align}}

\begin{document}

\unitlength = .8mm

\begin{titlepage}

\begin{center}

\hfill \\
\hfill \\
\vskip 1cm

\title{On The S-Matrix of Ising Field Theory \\ in Two Dimensions}

\author{Barak Gabai$^\spadesuit$, Xi Yin$^\spadesuit{}^\diamondsuit$}

\address{
$^\spadesuit$Jefferson Physical Laboratory, Harvard University, \\
Cambridge, MA 02138 USA
\\
$^\diamondsuit$Center for Theoretical Physics, Massachusetts Institute of Technology, \\
Cambridge, MA 02139 USA
}

\email{bgabai@g.harvard.edu, xiyin@g.harvard.edu}

\end{center}

\abstract{ We explore the analytic structure of the non-perturbative S-matrix in arguably the simplest family of massive non-integrable quantum field theories: the Ising field theory (IFT) in two dimensions, which may be viewed as the Ising CFT deformed by its two relevant operators, or equivalently, the scaling limit of the Ising model in a magnetic field. Our strategy is that of collider physics: we employ Hamiltonian truncation method (TFFSA) to extract the scattering phase of the lightest particles in the elastic regime, and combine it with S-matrix bootstrap methods based on unitarity and analyticity assumptions to determine the analytic continuation of the $2\to 2$ S-matrix element to the complex $s$-plane. Focusing primarily on the ``high temperature" regime in which the IFT interpolates between that of a weakly coupled massive fermion and the $E_8$ affine Toda theory, we will numerically determine 3-particle amplitudes, follow the evolution of poles and certain resonances of the S-matrix, and exclude the possibility of unknown wide resonances 
up to reasonably high energies.
}

\vfill

\end{titlepage}

\eject

\begingroup
\hypersetup{linkcolor=black}

\tableofcontents

\endgroup

\section{Introduction} 

The Ising field theory (IFT) in two spacetime dimensions, arguably the simplest non-integrable relativistic massive quantum field theory, is defined as the $c={1\over 2}$ Ising conformal field theory (ICFT) deformed by both the energy operator $\varepsilon$ and the spin field $\sigma$. The Euclidean action of IFT can be written as
\ie
S_{\rm IFT} = S_{\rm ICFT} + \int d^2x \left[  {m_f\over 2\pi}\, \varepsilon(x)  + h\, \sigma(x) \right].
\fe
A physical parameter of IFT is the dimensionless ratio of the couplings
\ie
\eta \equiv {m_f\over |h|^{8\over 15}},
\fe
loosely referred to as the ``temperature".\footnote{Note that the sign of $h$ is inconsequential due to the $\mathbb{Z}_2$ symmetry of the Ising CFT, whereas the sign of $m$ is physically significant.} In the ``high temperature" regime $\eta\gg 1$, the IFT is that of a weakly coupled massive fermion \cite{Zamolodchikov:2013ama}. In the opposite ``low temperature" regime $\eta\ll -1$, the IFT contains a tower of mesons that may be viewed as bound states of a pair of quarks tied with a flux string \cite{Fonseca:2006au}. At the special point $\eta=0$, the IFT is integrable and is well known to be equivalent to the $E_8$ affine Toda theory at the self-dual coupling \cite{Zamolodchikov:1989fp, Hollowood:1989cg, Braden:1989bu}.
In this paper we explore the non-perturbative S-matrix of IFT through a combination of Hamiltonian truncation and bootstrap methods, with the aim of determining the numerical values of the analytically continued S-matrix elements. 

More precisely, we adopt the truncated free fermion space approach (TFFSA) \cite{Yurov:1991my, Fonseca:2001dc} to compute the energy spectrum of IFT on the circle as a function of the radius $R$, and use Luscher's method \cite{Luscher:1985dn, Luscher:1986pf} to extract the elastic scattering phase of a pair of the lightest particles (of mass $m_1$) at center of mass energy $E$ below the inelastic threshold, namely $E<3m_1$ when there is only one stable particle, or $E<m_1+m_2$ when there is a second lightest particle of mass $m_2$($<2m_1$). We then use the elastic scattering data to determine the S-matrix on the complex $s$-plane ($s=E^2$), using unitarity at physical energies in the inelastic regime and the analyticity assumption that all poles on the first sheet are due to on-shell intermediate particles\footnote{The S-matrix elements under consideration are all free of anomalous thresholds.}. We will show that the analytically continued S-matrix elements can indeed be computed numerically with rigorous error bars, contingent upon numerical errors of TFFSA.

In particular, we will extract the on-shell three-point couplings of the first and second lightest particles (which is not directly accessible using Luscher's method as the energy lies in an unphysical regime), and identify the location of certain resonances (indicated by zeros of the S-matrix on the first sheet). Our results become less reliable numerically in the high energy regime. We will nonetheless be able to exclude the possibility of additional unknown resonances of sufficiently large width over a moderate energy range.

A quantitative relation between the finite size spectrum and resonance widths has been suggested by Luscher \cite{Luscher:1991cf}. The resonances of our consideration were previously studied in \cite{Delfino:1996xp,Pozsgay:2006wb}. Our main new ingredient is the Lemma of section \ref{sec:NAC}, which allows for bounding the error in determining the location of resonances, given sufficiently accurate low energy data. In a sense, our result illustrates the extent to which one can determine high energy physics through low energy ``experiments" in two spacetime dimensions, and it would be very interesting to establish analogous results for S-matrices in higher dimensions.

The paper is organized as follows. In section \ref{smatrixgen} we discuss the analytic structure of the $2\to 2$ S-matrix elements in two spacetime dimensions, and our method of determining the analytically continued S-matrix elements based on data in a certain range of physical energies. In section \ref{elasticampsec} we describe the determination of the scattering phases in IFT in the elastic regime using TFFSA. Our results for the analytically continued S-matrix elements of IFT are presented in section \ref{nacsec}. We discuss some physical implications and future applications and generalizations of our method in section \ref{discusssec}.

\section{S-matrix in two dimensions}
\label{smatrixgen}

\subsection{Analytic structure}

\begin{figure}[h!]
	\centering
	\def\svgwidth{11cm}
	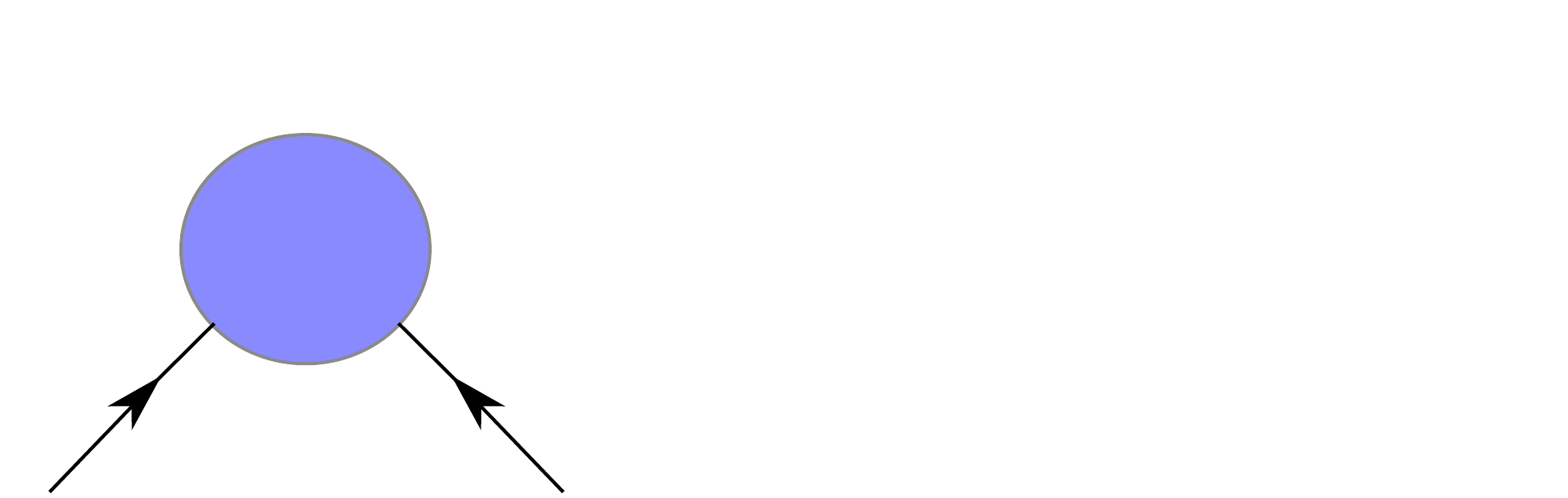
	\caption{In two-dimensional $2\to2$ scattering, energy-momentum conservation leaves the center of mass energy squared $s$ as the only kinematic invariant.}\label{fig:2dScatt}
\end{figure}

\begin{figure}[h!]
	\centering
	\def\svgwidth{16cm}
	\small{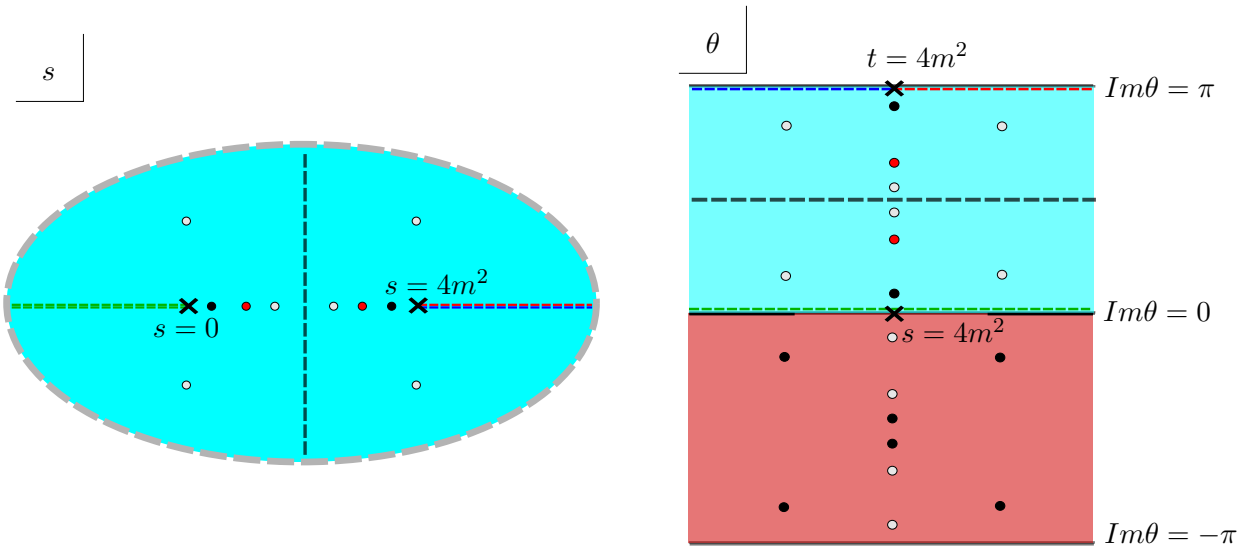}
	\caption{A typical set of poles (black and red dots) and zeros (grey dots) of a $2\to 2$ S-matrix element on the $s$-plane (left) and on the $\theta$-plane (right). The poles due to the self-coupling of the particle are represented by red dots. Blue and green dashed lines correspond to physical scattering in $s$ and $t$ channels respectively. On the physical strip of the $\theta$-plane (corresponding to the first sheet of the $s$-plane), zeros are allowed everywhere but they must appear in complex conjugate and crossing symmetric quadruplets (or pairs if they are on the imaginary line or on the crossing symmetric line), whereas poles are allowed only on for purely imaginary $\theta$ values. }
	\label{fig:sPlane}
\end{figure}

In this paper we focus on $2\to 2$ S-matrix elements of identical particles of mass $m$, 
with incoming momenta $p_1, p_2$ and outgoing momenta $-p_4, -p_3$, ordered according to
figure \ref{fig:2dScatt}. The Mandelstam variables are defined as \ie
s\equiv -(p_1+p_2)^2, ~~~ t\equiv -(p_2+p_3)^2=4m^2-s,~~~u\equiv -(p_1+p_3)^2 = 0.
\fe
In terms of the relative rapidity $\theta$ of particles 1 and 2, we can write
\ie
s = 4m^2 \cosh^2{\theta\over 2}.
\fe
Note that the map from $s$ to $\theta$ is not single-valued, and the choice of branch will be determined through the analytic continuation procedure (see figure \ref{fig:sPlane}).

The $2\to 2$ S-matrix element, denoted simply $S(s)$, is an analytic function on the complex $s$-plane (``first sheet") away from the branch cuts along $(-\infty, 0] \cup [4m^2, \infty)$, the $s$-channel pole at $s=m^2$ and the $t$-channel pole at $s=3m^2$, as well as other possible poles corresponding to on-shell particles in the $s$ or $t$ channel.

We will further assume that the particle of mass $m$ is the lightest particle in the theory. Under this assumption there will be no further singularities on the $s$-plane corresponding to anomalous thresholds \cite{Eden:1966dnq, Goddard:1969ci}. Resonances, on the other hand, correspond to poles on the ``second sheet" of $S(s)$ analytically continued through the branch cut of $2$ particle states of the lightest particle.

Crossing symmetry amounts to 
\ie
S(s) = S(4m^2-s).
\fe
The physical $2\to 2$ scattering amplitude is given by the value of $S(s)$ approaching the right branch cut from above, namely
\ie
\lim_{\epsilon \to 0^+} S(s+i \epsilon),~~~ s \in [4m^2, \infty).
\fe
For physical energies in the $s$-channel, $S(s)$ is subject to the unitarity constraint
\ie
|S(s)|\leq 1,~~~ s\in [4m^2, \infty)
\fe
In particular, below the inelastic threshold, namely for $4m^2 \leq s\leq {\rm min}((m+m_2)^2, 9m^2)$, where $m_2$ is the mass of the second lightest particle (if it exists), we have $|S(s)|=1$.

Furthermore, $S(s)$ is subject to the real analyticity condition
\ie
S(s) = (S(s^*))^* \equiv S^*(s),
\fe
where $S(s^*)$ is defined as the value of $S(s')$ at $s'=s^*$ on the first sheet. It follows that the value of $S(s)$ approaching the branch cut $[4m^2,\infty)$ from below is the complex conjugate of the physical amplitude.

On the complex rapidity $\theta$-plane, the branch cuts are located at $\theta \in (-\infty, -\theta_0] \cup [\theta_0, \infty)$ along the real $\theta$ axis and their translations by integer multiples of $\pi i$, where $\theta_0$ is the rapidity value that corresponds to the inelastic threshold. The pole at $s=m^2$ and $3m^2$ translate to poles at $\theta = {2\pi i\over 3}$ and ${\pi i\over 3}$, as well as their images under shifts by integer multiples of $2\pi i$. Physical amplitudes are given by
\ie
\lim_{\epsilon\to 0^+} S(\theta + i\epsilon), ~~~\theta\in\mathbb{R}_{\geq 0}.
\fe
The first sheet of the complex $s$-plane is mapped to the ``physical strip" $0<{\rm Im}\,\theta< \pi$. The second sheet corresponds to $-\pi < {\rm Im}\,\theta<0$. The crossing symmetry is expressed in terms of $S(\theta)$ as
\ie
S(\theta) = S(\pi i -\theta).
\fe
The real analyticity condition can be expressed as
\ie
S(\theta) = (S(-\theta^*))^* \equiv S^*(-\theta).
\fe
This is because the regions above and below the physical cut $s\in [4m^2, \infty)$ on the $s$-plane, related by complex conjugation, are mapped to the regions above the positive and negative real $\theta$ axis on the $\theta$-plane, which are related by complex conjugation combined with a sign flip (see figure \ref{fig:sPlane}).

The unitarity constraint can be expressed as
\ie\label{unitff}
\lim_{\epsilon\to 0^+}S(\theta + i \epsilon) S(-\theta+i\epsilon) = f(\theta),
\fe
where $f(\theta)$ is a real-valued function for $\theta\in \mathbb{R}$ that obeys $f(\theta)=1$ for $|\theta|< \theta_0$ and $f(\theta)\leq 1$ for $|\theta|\geq \theta_0$. Since $S(\theta)S^*(\theta)=1$ in the elastic regime, we can analytically continue this relation to the whole physical strip, and write 
\ie
S(\theta) = {1\over S^*(\theta)} 
={1\over S(-\theta)}.
\fe
It follows that poles in the physical strip (first sheet) correspond to zeros on the second sheet and vice versa (see figure \ref{fig:sPlane}).

It is generally expected from causality \cite{Eden:1966dnq, Camanho:2014apa} that all poles on the $\theta$-plane away from the above mentioned branch cuts are either due to intermediate on-shell particles (poles on the physical strip) or resonances (poles on the second sheet). 
It is further expected that in a local quantum field theory $S(s)$ is bounded near infinity on the $s$-plane
\cite{Jaffe:1966an, Epstein:1969bg, Paulos:2016but}.\footnote{The usual statement is that $S(s)$ is polynomial bounded at large complex $s$. Together with the unitarity bound along the physical branch cut, a theorem of Fuchs \cite{Fuchs} implies that $S(s)$ is in fact bounded at large complex $s$.} Under these assumptions, we can write $S(\theta)$ in the form
\ie\label{sgenstr}
S(\theta) = S_{\rm CDD}(\theta) \exp\left[- \int_{-\infty}^\infty {d\theta'\over 2\pi i} {\log f(\theta')\over \sinh(\theta-\theta'+i\epsilon)} \right],
\fe
where $f(\theta)$ is the function appearing in (\ref{unitff}). $S_{\rm CDD}(\theta)$, known as the Castillejo-Dalitz-Dyson (CDD) factor, is meromorphic on the physical strip and is a pure phase for real $\theta$. 

If one further assumes that there is no essential singularity at infinity, then the CDD factor can be written as 
\ie \label{eq:CDDfacs}
S_{\rm CDD}(\theta) = \pm \prod_j {i \sin \A_j + \sinh\theta \over i \sin \A_j - \sinh \theta}.
\fe
The $\A_j$'s with ${\rm Re}(\A_j)\in (0, \pi)$ give rise to poles in the physical strip, which we refer to as ``CDD poles", whereas those with ${\rm Re}(\A_j)\in (-\pi, 0)$ give rise to poles on the second sheet, or zeros in the physical strip, which we refer to as ``CDD resonances" or ``CDD zeros". The CDD poles are due to intermediate on-shell particles, and are expected to lie on the imaginary $\theta$-axis (absent anomalous thresholds). The CDD zeros could be anywhere in the physical strip, and would generally come in quadruples with respect to the reflection $\theta\to -\theta^*$ and with respect to crossing transformation $\theta\to \pi i-\theta$, unless either ${\rm Re}\,\theta=0$ or ${\rm Im}\,\theta={\pi\over 2}$.

In the rest of this paper, while we will be guided by (\ref{eq:CDDfacs}) in producing the numerical approximation of $S(\theta)$, the method of bounding the error in section \ref{sec:NAC} does not require the assumption of absence of essential singularity at infinity.

\subsection{Numerical analytic continuation and bounds on error}
\label{sec:NAC}

\begin{figure}[h]
	\centering
	\def\svgwidth{17cm}
	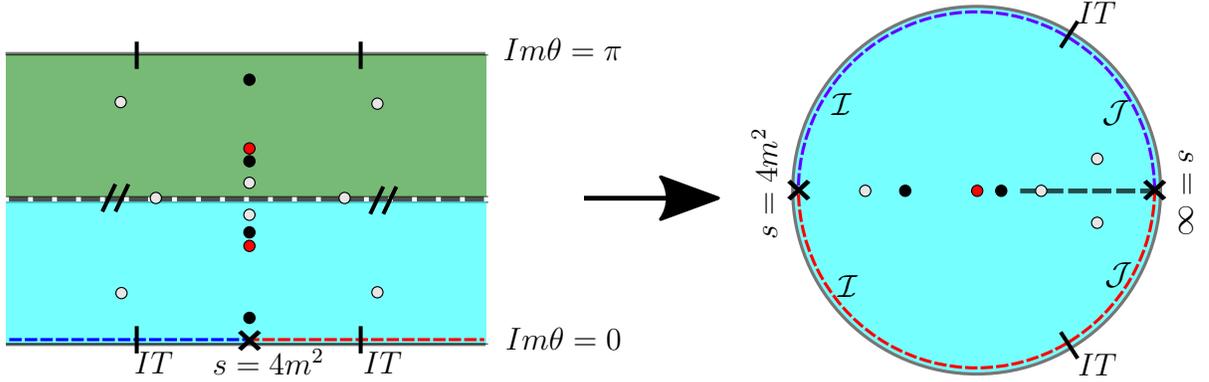
	\caption{Mapping from the $\theta$-plane (left) to the $z$-disc (right). The red dots, representing the self-coupling pole, are mapped to the center of the $z$-disc. Other typical poles are represented by black dots, whereas typical zeros are represented by grey dots. The particle production threshold is indicated by the black stubs. }\label{fig:zDisc}
\end{figure}

To proceed, it will be convenient to introduce the $z$ variable
\ie\label{zmap}
z = {\sinh\theta - {\sqrt{3}\over 2} i\over \sinh\theta + {\sqrt{3}\over 2} i} = {\sqrt{s(4m^2-s)} - \sqrt{3} m^2\over \sqrt{s(4m^2-s)} + \sqrt{3} m^2}.
\fe
(\ref{zmap}) maps half of the physical strip, namely the domain $0\leq {\rm Im}\, \theta \leq {\pi\over 2}$, to the unit disc $|z|\leq 1$. The physical amplitude is given by the value of $S(z)$ at the boundary of the unit $z$-disc. We will denote by ${\cal I}$ the arc on the boundary of the disc that correspond to the image of $[-\theta_0, \theta_0]$, where $S(z)$ is a phase, and by ${\cal J}$ the complement of ${\cal I}$ on the boundary of the disc that correspond to energies above the inelastic threshold where $|S(z)|\leq 1$ (see figure \ref{fig:zDisc}).

The $t$-channel pole at $\theta={\pi i\over 3}$, due to the exchange of a particle that is identical to the external particles, is mapped to $z=0$. Any other poles of $S(z)$ on the unit $z$-disc would correspond to the exchange of another (stable) particle, whose mass is between $m$ and $2m$ (since we have assumed that the external particle of mass $m$ is the lightest). Such a pole corresponds to a CDD factor of the form
\ie
{1- z_* z\over z_*-z},
\fe
where $z_*$ lies on the interval $(-1,7-4\sqrt{3})$ (corresponding to purely imaginary values of $\theta$).

Resonances, on the other hand, correspond to zeros of $S(z)$ in the unit disc. They come in complex conjugate pairs due to the relation $S(z) = (S(z^*))^*$, unless they lie on the real $z$-axis.

The function
\ie\label{gfunc}
g(z) \equiv - S(z) z \prod_j {z_j-z\over 1-z_j z},
\fe
where the product over $j$ ranges over all CDD poles other than the one at $z=0$, is analytic in the interior of the unit $z$ disc, obeys $|g(z)|\leq 1$ on the boundary of the disc, and by assumption is bounded near $z=1$ in the interior of the disc. Note that we will not need to assume the absence of essential singularity at $z=1$, allowing for instance the possibility of gravitational dressing factors in the S-matrix \cite{Dubovsky:2012wk, Paulos:2016but}.

Our objective is to determine $g(z)$ from its values on the arc ${\cal I}$, namely the elastic scattering phases that will be computed independently in the next section using TFFSA and Luscher's method. 
For this, we make use of the following 

\noindent{\bf Lemma:} Given any function $F(z)$ that is analytic and bounded in the interior of the unit disc, and continuous at the boundary,\footnote{The assumption of continuity on the boundary can be dropped if we replace the integrals over the boundary arcs in (\ref{lemmabound}) and (\ref{bzforms}) by the suprema of the corresponding integrals over nearby curves in the interior of the disc, but this will not be necessary for our application.} and an open subset ${\cal I}$ of the boundary (a collection of arcs), we have the inequality
\ie\label{lemmabound}
|F(z)| \leq B_z^{1- {1\over 2\pi}|\tau_z^{-1}({\cal I})|} \exp\left[ \int_{\tau_z^{-1}({\cal I})} {du\over 2\pi i u}\log |F(\tau_z(u))| \right]
\fe
for $z$ in the interior of the disc. Here $|\tau_z^{-1}({\cal I})|$ is the total length of $\tau_z^{-1}({\cal I})$, where $\tau_z$ is the M\"obius map 
\ie\label{mobius}
\tau_z(w) = {- \bar z z + z(w+i) + iw\over -\bar z + \bar z(z+i) w + i}
\fe
that takes the unit disc to itself while taking 0 to $z$, and $B_z$ is defined as
\ie\label{bzforms}
B_z = {1\over |\tau_z^{-1}({\cal J})|} \int_{\tau_z^{-1}({\cal J})} {dw\over i w } |F(\tau_z(w))|,
\fe
where ${\cal J}$ is the complement of ${\cal I}$ in the unit circle. 

The proof of the above lemma is given in Appendix \ref{app:strongerLemma}. 

For our application, $F(z)$ will be the difference between the function $g(z)$ defined in (\ref{gfunc}) and a trial function $\widetilde g(z)$ with similar properties (e.g. $|\widetilde g(z)|\leq 1$ on the unit circle) that closely approximates $g(z)$ in the elastic regime (the arc ${\cal I}$). The lemma then places a bound on the ``error" $|g(z)-\widetilde g(z)|$ in the interior of the disc. In this case, $B_z\leq 2$, and we will simply replace $B_z$ by 2 in practical evaluations of the error bound.
We now describe two practical methods of constructing such numerical approximations of $g(z)$.
\bigskip

\noindent{\bf Method I.} We can express $g(z)$ in terms of its boundary values using the contour representation
\ie
g(z) = \oint_C {dw\over 2\pi i \chi(z)} {\chi(w)\over w-z} g(w)
\fe
where $\chi(w)$ is a bounded analytic function in the interior of the disc. If $\chi(w)/\chi(z)$ is sufficiently small in the ``inelastic region" $w\in {\cal J}$, one may simply drop the contribution from the ``inelastic region" in the contour integral, and approximate $g(z)$ with
\ie\label{gbarapprox}
\widetilde{g}(z) &= \int_{{\cal I}} {dw\over 2\pi i\chi(z)} {\chi(w)\over w-z} g(w).
\fe
In other words, given a choice of the function $\chi(z)$, the approximation $\widetilde{g}(z)$ is entirely determined by the elastic scattering phase. It is evident that the error is bounded by
\ie
\left| g(z) - \widetilde{g}(z) \right|  \leq \int_{{\cal J}} {dw\over 2\pi i w |\chi(z)|} {|\chi(w)| \over |w-z|} .
\fe
A class of candidate $\chi$ function is given by (\ref{chiformFull}), (\ref{hformFull}), although such functions are typically highly oscillatory on the boundary of the disc and may not be suitable for numerically evaluating (\ref{gbarapprox}). More generally, an admissible $\chi(z)$ can be constructed as
\ie\label{chipnz}
\chi(z) = 1 - P_N(z),~~~ P_N(z) = \sum_{n=1}^N a_n z^n, 
\fe
such that $P_N(z)$ approximates 1 on the arc ${\cal J}$. The latter is always possible because the set of functions $\{z^n|n\geq 1\}$ is a complete basis for $L^2$ functions on ${\cal J}$ (although not on the entire circle), by virtue of Beurling-Malliavin theorem. One may attempt to choose such coefficients $a_n$ to minimize the gradient of $\chi(z)$ along the unit circle to make numerical integration more feasible.

\bigskip

\noindent{\bf Method II.} Alternatively, one may make some assumptions on the number of (resonance) zeros of $g(z)$, or generally the structure of the factor $S_{\rm CDD}(\theta)$ in (\ref{sgenstr}), and directly fit their positions and the function $f(\theta)$ against the elastic scattering phase to obtain an approximation $\widetilde{g}(z)$. While one may in principle miss some resonance zeros, by our lemma, as long as $\widetilde{g}(z)$ approximates $g(z)$ in the elastic regime, its error anywhere in the interior of the disc can be bounded rigorously (though not uniformly). 
In particular, it allows us to rigorously rule out the possibility of additional resonances in certain regions of the disc. 

In terms of the $z$ variable, we can write (\ref{sgenstr}) as
\ie\label{soverscdd}
S(z)= S_{\rm CDD}(z) \exp\left[ \int\limits_{0}^{\phi_0}\frac{d\phi}{2\pi} \left( {e^{i\phi}+z\over e^{i\phi}-z} + {e^{-i\phi}+z\over e^{-i\phi}-z} \right)\log \mathfrak{f}(\phi) \right] ,
\fe
where $\mathfrak{f}(\phi)$ is $f(\theta)$ evaluated at $z=e^{i\phi}$. $-\log\mathfrak{f}(\phi)$ is expected to be a positive even function in the range $\phi\in (-\phi_0, \phi_0)$, corresponding to energies above the particle production threshold.

Our approximation to $g(z)$ takes the form
\ie\label{phasfit}
\widetilde{g}(z)= \left( \prod_{k=1}^M {z_k'-z\over 1-z_k' z}\right) \exp\left[ -\int\limits_{0}^{\phi_0}\frac{d\phi}{2\pi} \left( {e^{i\phi}+z\over e^{i\phi}-z} + {e^{-i\phi}+z\over e^{-i\phi}-z} \right) Q(\phi/\phi_0) \right]  ,
\fe
where $Q(x)$ is taken to be a sum of squares of polynomials, constrained by $Q(1)=0$. The ``trial zeros" $z_1',\cdots, z_M'$ as well as $Q(x)$ are chosen by fitting the phase of $\widetilde g(z)$ to that of $g(z)$ on the elastic arc ${\cal I}$.

\bigskip
Application of both methods will be illustrated in section \ref{nacsec}. We find method II more feasible in practice, as in many cases the scattering phase can be reliably extracted from TFFSA only on a set of disjoint arcs in the elastic regime. Nonetheless, the error bound given by the lemma is typically quite small away from the inelastic arc ${\cal J}$, when reasonable assumptions on resonance zeros are made. A priori, zeros of $S(z)$ can appear anywhere on the $z$-disc, so long as they come in complex conjugate pair when they are away from the real line. In IFT, as we increase $\eta$ from zero to positive values, the 5 heavier particles of $E_8$ theory are expected to turn into complex resonances. The simplest possibility is that the only other zeros of $S(z)$ appear on the real interval $(-1,1)$. Our analysis will be consistent with this scenario in that, by numerically fitting (\ref{phasfit}) with $z_k'$ constrained to lie on the real interval $(-1,1)$, we will be able to rigorously bound the error of our approximation to $S(z)$ on the $z$-disc, and exclude the possibility of additional complex zeros outside of a small neighborhood of the inelastic arc ${\cal J}$. 

Let us comment that the existence of analytic continuation to the interior of the $z$-disc subject to the unitarity bounds on the boundary and the analytic assumptions imposes highly nontrivial constraints on the elastic scattering data themselves. Numerical errors in the elastic data (as obtained from TFFSA) could in principle lead to different fits of $S(z)$ with incompatible (e.g. mutually exclusive) error bounds.

Note that the error bound becomes loose near the inelastic arc on the boundary of the $z$-disc. Indeed, based on effective field theory one expects that the low energy scattering data is insensitive to a high energy resonance of very small width.

\section{Elastic amplitudes of Ising field theory}
\label{elasticampsec}

\subsection{TFFSA for Ising field theory}

To compute the energy spectrum of Ising field theory on a cylinder of radius $R$, we write the Hamiltonian as
\ie \label{eq:IFTHam}
H = H_0 + h \int_0^{2\pi R} dx\, \sigma(x),
\fe
where $H_0$ is the Hamiltonian of a free Majorana fermion field of mass $|m_f|$ that results from the Ising CFT deformed by the energy operator. The free fermion Hilbert space ${\cal H}^{\rm free}$ and the spectrum of $H_0$, however, depend on the sign of the deformation parameter $m_f$. ${\cal H}^{\rm free}$ is the direct sum of Neveu-Schwarz (NS) and Ramond (R) sectors,
\ie
{\cal H}^{\rm free} = {\cal H}_{\rm NS} \oplus {\cal H}_R, 
\fe
where ${\cal H}_{\rm NS}$ is spanned by states of the form
\ie
|n_1, \cdots, n_N\rangle_{\rm NS},~~~ n_i \in \mathbb{Z}+{1\over 2},
\fe
where $N$ is an even positive integer, and $n_i$ are distinct momentum quantum numbers (in units of $1/R$) of the $N$ fermions. The R sector, on the other hand, is spanned by states of the form
\ie
|n_1, \cdots, n_N\rangle_{\rm R},~~~ n_i \in \mathbb{Z},
\fe
where $N$ is odd for the ``high temperature" theory with $m_f>0$, and even for the ``low temperature" theory with $m_f<0$. A fermion mode labeled by $n$ carries momentum $p_n$ and energy $\omega_n$,
\ie
p_n = {n/ R}, ~~~ \omega_n = \sqrt{m_f^2 + (n/R)^2}.
\fe
The ground state energy of NS and R sector are denoted $E_0^{\rm NS}(R)$ and $E_0^{\rm R}(R)$ respectively, with their difference given by
\ie\label{nsrdiff}
E_0^{\rm NS}(R)-E_0^{\rm R}(R) = \int_{-\infty}^\infty {dp\over 2\pi} \log{1-e^{-2\pi R E}\over 1+e^{-2\pi R E}},
\fe
where we write $E\equiv \sqrt{p^2+m_f^2}$. Note that in the large $R$ limit, (\ref{nsrdiff}) is of order $e^{-|m_f| R}$.

The spin field $\sigma$ has only nonzero matrix elements (form factors) between NS and R sector states, given by \cite{Fonseca:2001dc}\footnote{In \cite{Fonseca:2001dc} $R$ denotes the circumferences of the circle, whereas here we use $R$ to denote the radius, hence a factor $2\pi$ difference.}
\ie \label{eq:FSTrans}
{}_{\rm NS}\langle k_1, \cdots, k_M|\sigma(0)|n_1, \cdots, n_N\rangle_{\rm R} = e^{s(R)} \prod_{i=1}^N {e^{\kappa(R, p_{n_i})}\over \sqrt{2\pi R E_{n_i}}} \prod_{j=1}^M {e^{-\kappa(R, p_{k_j})}\over \sqrt{2\pi R E_{k_j}}} F(k_1,\cdots, k_M|n_1,\cdots, n_N),
\fe
where
\ie
F(\{k_i\}|\{n_{i'}\}) = i^{\left[\frac{M+N}{2}\right]} \bar \sigma \prod_{i<j} {p_{k_i} - p_{k_j}\over E_{k_i}+E_{k_j}} \prod_{i'<j'} {p_{n_{i'}}-p_{n_{j'}}\over E_{n_{i'}}+E_{n_{j'}}} \prod_{i,j'} {E_{k_i} + E_{n_{j'}}\over p_{k_i} - p_{n_{j'}}},
\fe
\ie
\bar \sigma = \bar s |m_f|^{1/8}\,,\qquad \bar s = 2^{1/12} e^{-1/8} A_G^{3/2} = 1.35783834\cdots,
\fe
with $A_G$ being Glaisher’s constant and the functions $s(R)$ and $\kappa(R, p)$ are given by
\ie{}
& s(R) =  { R^2\over 2} \int_{-\infty}^\infty {dp dp' \over E E'} {p p'\over \sinh(2\pi R E) \sinh(2\pi R E')} \log {E+ E'\over |p-p'|},
\\
& \kappa(R, p) = m_f^2 \int_{-\infty}^\infty {dp'\over 2\pi E'} {1\over EE'-pp'} \log {1-e^{-2\pi R E'}\over 1+ e^{-2\pi R E'}},
\fe
where we again wrote $E\equiv \sqrt{p^2+m_f^2}$ and similarly for $E'$. Note that both $s(R)$ and $\kappa(R, p)$ vanish in the $R\to \infty$ limit. 

\begin{figure}[h]
\centering
\includegraphics[width=0.5
\linewidth]{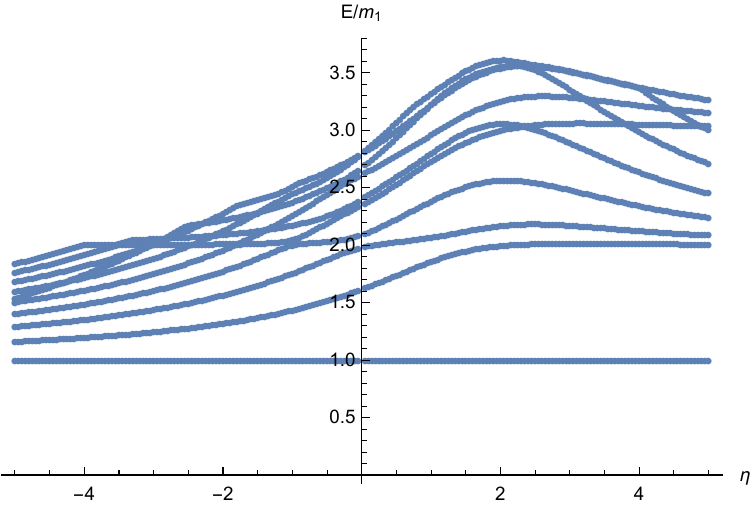}
\caption{Level 19 result for energy levels as a function of $\eta$. The cylinder circumference is taken to be $2\pi R = 6\eta/m_f$, where $m_f$ is the free fermion mass. Here we have subtracted the ground state energy and normalized the energy levels with respect to the mass $m_1$ of the lightest particle. The parts of the curves for $E/m_1$ between 1 and 2 correspond to stable particles, whereas those with $E/m_1>2$ represent scattering states of two or more particles.
}
\label{fig:etaElvlsLvl19}
\end{figure}

We will restrict to the sector of the Hilbert space with zero total momentum along the circle. Starting from the free massive fermion Hilbert space, including both NS and R sectors, we will truncate to the finite dimensional subspace spanned by states involving any number of left or right moving fermions, such that the total right moving momentum is no more than $L/R$. $L$ will be referred to as the ``truncation level". We then evaluate the matrix elements of the IFT Hamiltonian for a basis of the truncated Hilbert space, and numerically diagonalize the resulting finite dimensional Hamiltonian matrix to obtain the energy spectrum. This is known as the TFFSA \cite{Yurov:1991my}. The error of such an approximation to the energy levels is expected to be suppressed by inverse powers of the truncation level $L$, as a consequence of asymptotic decoupling between low and high energy states \cite{Giokas:2011ix, Hogervorst:2014rta}. 

A typical set of resulting energy spectrum obtained from TFFSA is shown in figure \ref{fig:etaElvlsLvl19}. In this example we have taken the truncation level $L=19$, fixing a suitable choice of the circle radius $R$ in units of $|h|^{-{8\over 15}}$, and scanned over a range of $\eta$. In the plot we displayed the first 10 energy levels after subtracting the ground state energy, and normalized in units of the first excitation energy (identified with the mass $m_1$ of the lightest particle). The energy levels below $2m_1$ correspond to stable particles, whereas those above $2m_1$ generically correspond to scattering states.

To minimize the error due to level truncation, we can perform an extrapolation to $L=\infty$ by fitting the finite level data to
\begin{equation}
\label{eq:extrapToInf}
E^{(L)}(R) = E^{(\infty)}(R) + \alpha(R) \,L^{-\beta(R)},
\end{equation}
following \cite{Fonseca:2001dc}. In determining the mass of the stable particles, we first extrapolate the excitation energies to $L=\infty$ at fixed radii $R$, and then extrapolate to infinite radius using an exponential fit.

As for the scattering states in the elastic regime, we work at a given truncation level $L$ and extract the scattering phase using Luscher's method as explained in the next section, for a range of radii. We then extrapolate the resulting graph for the scattering phase as a function of energy to $L=\infty$ by a power law fit analogous to (\ref{eq:extrapToInf}).

\begin{figure}[h!]
	\centering
	\includegraphics[width=0.45\linewidth]{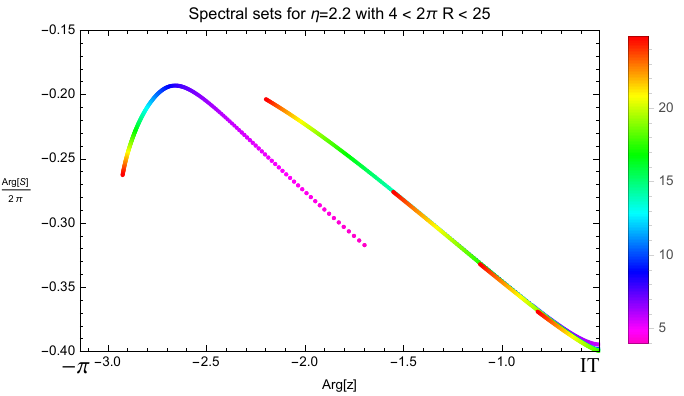}~~~~
	\includegraphics[width=0.45\linewidth]{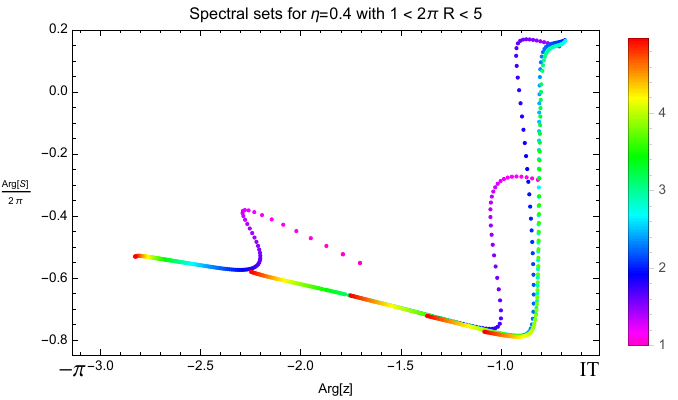}
	\caption{A demonstration of finite size effects in the spectral sets, computed using TFFSA at truncation level $L=22$ for two sample values of $\eta$. The vertical axis is the approximation to the elastic scattering phase extracted from the Bohr-Sommerfeld quantization condition, and the horizontal axis is $\arg(z)$ (related to the energy via (\ref{zmap})). The value of the cylinder radius $R$ is indicated by the color scheme, from red (large radii) to violet (small radii). 
	For $\eta=2.2$, in the range $-2.7<\arg{z}<-1.6$, the spectral set of the lowest 2-particle state at small radii and that of the second lowest 2-particle state at larger radii do not overlap, due to significant finite size effect in the former. To handle this problem, we discard a posterior the spectral set data for the lowest 2-particle state at $2\pi R\lessapprox10$. The finite size effects are significantly smaller for the higher 2-particle states, as seen from their approximate overlaps. For $\eta=0.4$, finite size effects plague higher level 2-particle states as well for $2\pi R\lessapprox 3$, again in a clearly visible manner. }
	\label{fig:FSeffectsOnPhase}
\end{figure}

\begin{figure}[h!]
	\centering
	\includegraphics[width=.75\linewidth]{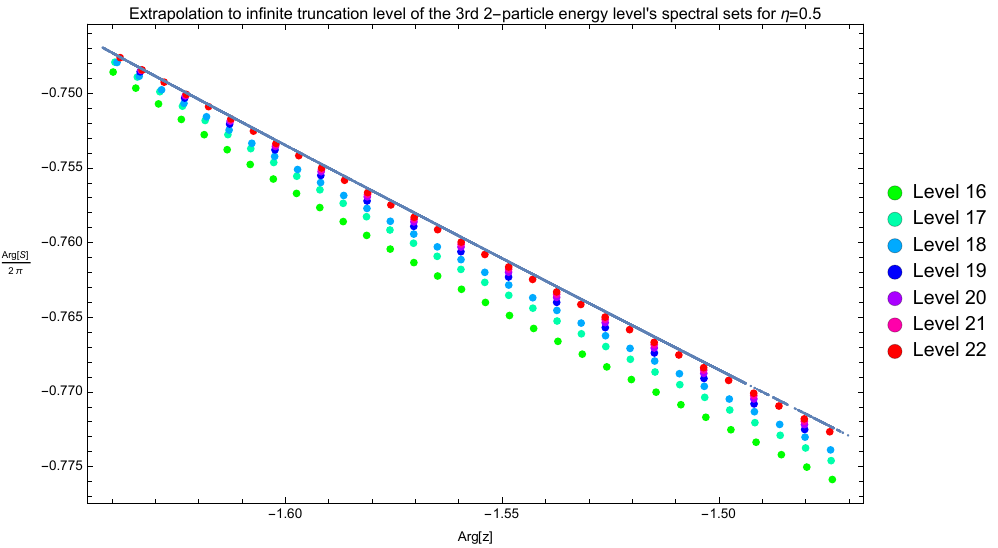}
	\caption{A demonstration of the convergence of the spectral set with respect to increasing truncation level $L$. Here we consider the spectral set extracted from a fixed 2-particle energy level by sampling over a range of radii, at $\eta=0.5$. 
We interpolate the spectral set within the given energy level, for $L$ ranging from 16 to 22, and extrapolate the phase values at each $\arg(z)$ to $L=\infty$ by fitting to a power law convergence, resulting in the blue curve. }
	\label{fig:extrapAnElvl}
\end{figure}

\subsection{Extracting scattering phases}
\label{sec:extracphase}

\begin{figure}[h!]
	\centering
	\includegraphics[width=.45
	\linewidth]{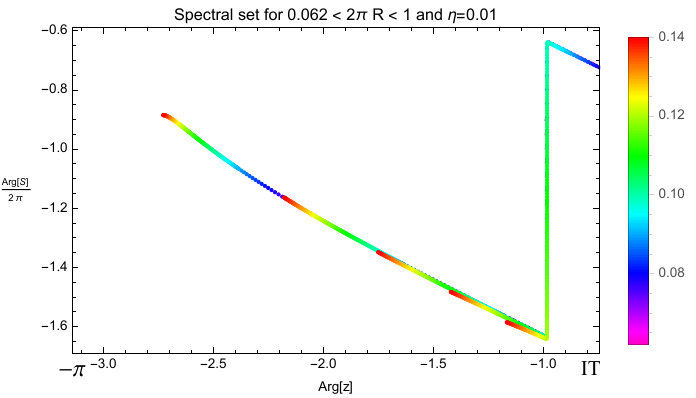}~~~~~
	\includegraphics[width=.45
	\linewidth]{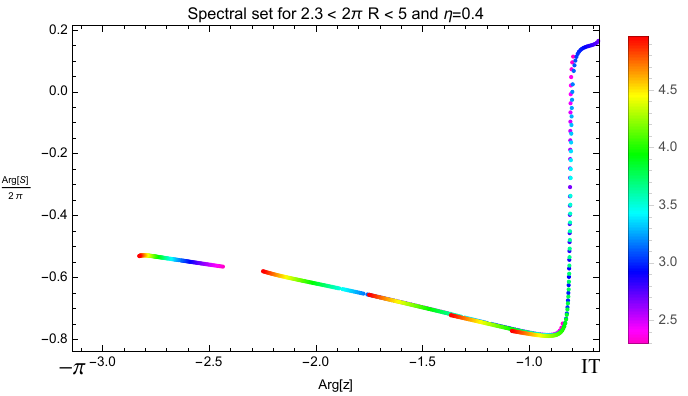}
	\centering
	\includegraphics[width=.45
	\linewidth]{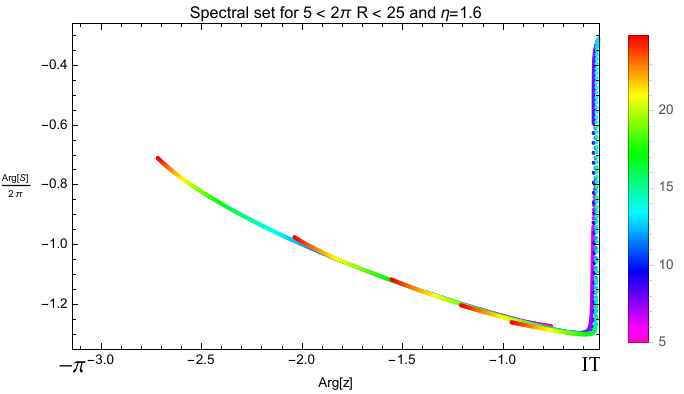}~~~~~
	\includegraphics[width=.45
	\linewidth]{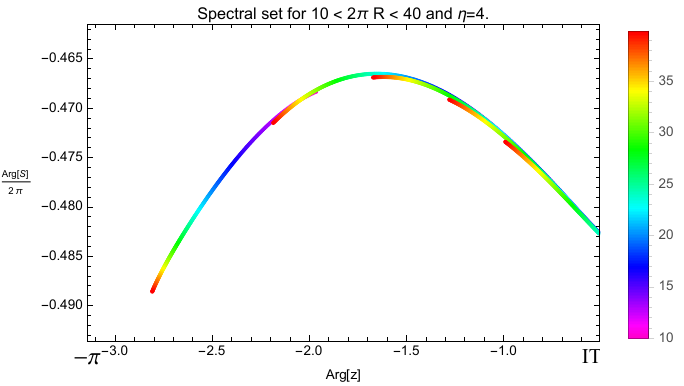}
	\caption{The spectral sets for the scattering phase of a pair of lightest particles as a function of $\arg(z)$ in the elastic regime (from $-\pi$ to the inelastic threshold), computed at truncation level $L=22$ over a suitable range of the cylinder radius $R$ (in units of $1/m_f$), for $\eta=0.01$, $0.4$, $1.6$, and $4$ respectively.  
}
	\label{fig:phaseByLevel}
\end{figure}

\begin{figure}[h!]
		\includegraphics[width=.45\linewidth]{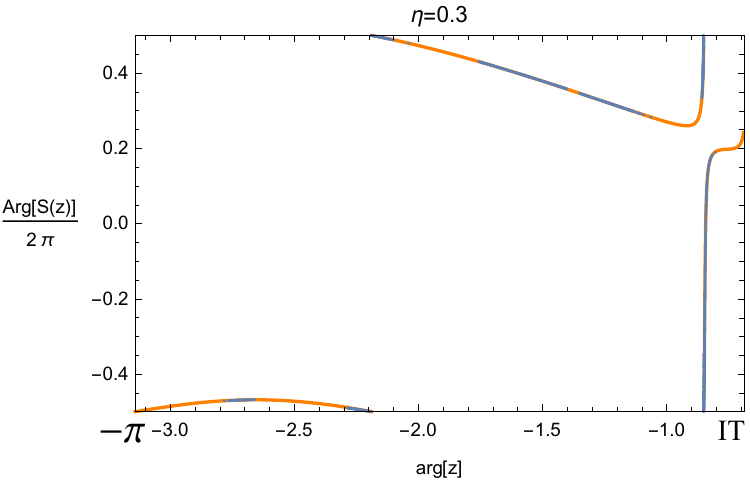}~~~~~
		\includegraphics[width=.45\linewidth]{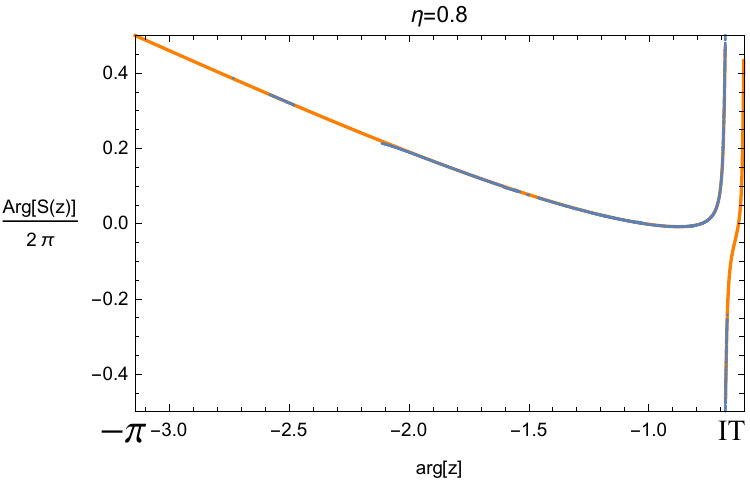}\\
		\includegraphics[width=.45\linewidth]{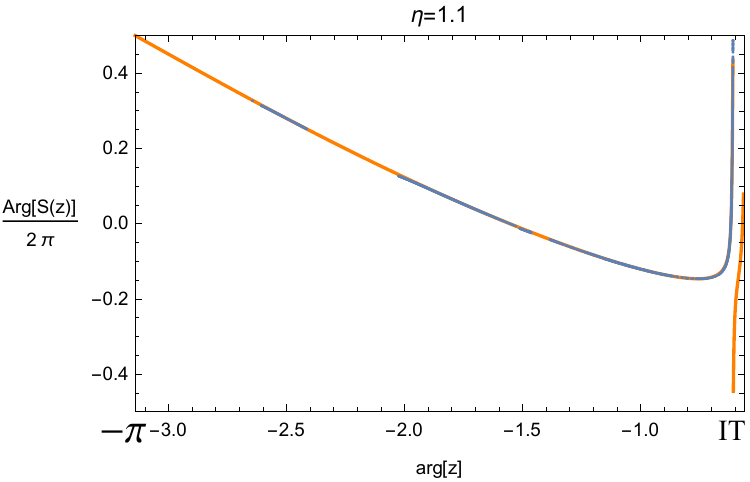}~~~~~
		\includegraphics[width=.45\linewidth]{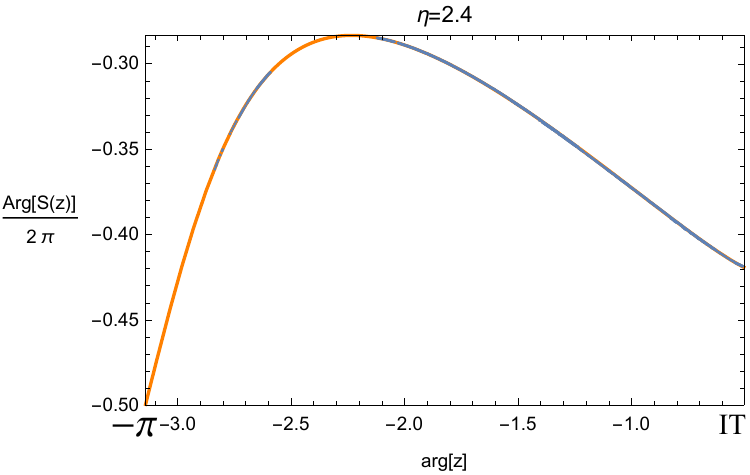}
	\caption{The scattering phase as a function of $\arg(z)$ in the elastic regime, obtained from TFFSA via Luscher's method, is shown in blue. The orange curve represents the fit to an analytic function built out of a presumed number of CDD zeros and an exponential factor in the form of (\ref{phasfit}), multiplied by the known CDD pole factors (determined by the masses of stable particles).}
	\label{fig:eta_infLvlFit}
\end{figure}

Focusing on the energy range of elastic scattering states of two of the lightest particle with zero total momentum, we define the spectral set
\ie \label{eq:specstets}
{\cal S} = \left\{ \left.(E_n, n - R p_n) \right|\, 2m_1<E_n<{\rm min}(m_1+m_2, 3m_1) \right\}.
\fe
Here the energies $E_n$ are labeled by the level $n$ in increasing order, with the understanding that for sufficiently large radius $R$, $E_n$ can be thought of as the scattering state of a pair of particles carrying momentum $p_n$ and $-p_n$ on the circle, with
\ie
E_n = 2 \sqrt{p_n^2 + m_1^2},
\fe
such that $p_n$ is subject to the Bohr-Sommerfeld quantization condition
\ie
\label{eq:BS-quant}
2\pi R p_n + \varphi(E_n) = 2\pi n,
\fe
where $\varphi(E)$ is the elastic scattering phase at center of mass energy $E$. Thus, for sufficiently large $R$, the spectral set ${\cal S}$ is expected to lie on the graph of the function ${1\over 2\pi} \varphi(E)$. Examples of finite size effects are shown in Figure \ref{fig:FSeffectsOnPhase}. On the other hand, errors due to the level truncation in TFFSA becomes worse at large $R$. The convergence with increasing truncation level is demonstrated in Figure \ref{fig:extrapAnElvl}.

In practice, it is typically sufficient to take the union of spectral sets collected over suitable moderate ranges of $R$. The smallness of finite size effects is justified a posteriori by observing the approximate overlap of spectral sets for different energy levels of 2-particle states, as demonstrated in Figure \ref{fig:phaseByLevel}.

\section{The analytically continued S-matrix of Ising field theory}
\label{nacsec}

\subsection{Numerical determination of $S(z)$ on the disc using method II}

We now present results on the $2\to 2$ scattering amplitude $S(z)$ of the lightest particle in IFT, analytically continued into the interior of the complex $z$-disc (or equivalently, to the physical strip in the $\theta$-plane). We adopt method II of section \ref{sec:NAC}, in which we begin by making simple assumptions on the number of CDD/resonance zeros of $S(z)$ on the disc, and then fit both the CDD zero locations and the function $\mathfrak{f}(\phi)$ in (\ref{soverscdd}) against the elastic scattering phase computed using TFFSA via Luscher's method.

Our data for the elastic phase is subject to two types of errors: those due to level truncation on the Hamiltonian in TFFSA, and finite size effects on the circle of radius $R$. The former is handled by an extrapolation to infinite truncation level using (\ref{eq:extrapToInf}). The finite size effects, which corrects both the mass of the particles and the quantization condition (\ref{eq:BS-quant}), is exponentially suppressed at sufficiently large $R$. As the TFFSA data are collected over a range of $R$, the finite size effects are handled by imposing a suitable lower bound on $R$, as illustrated in section \ref{sec:extracphase}. Consequently, the reliable data for the elastic phase may only cover a collection of disjoint arcs ${\cal I}'$ within the elastic arc ${\cal I}$ on the boundary of the $z$-disc. In this case, we simply replace ${\cal I}$ by ${\cal I}'$ in applying the method of section \ref{sec:NAC}.

\begin{figure}[h!]
	\centering
	\includegraphics[width=0.45\linewidth]{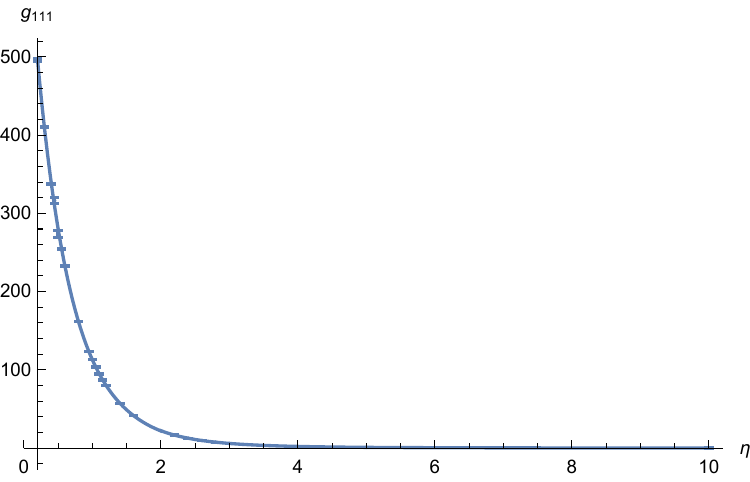}
	\includegraphics[width=0.45\linewidth]{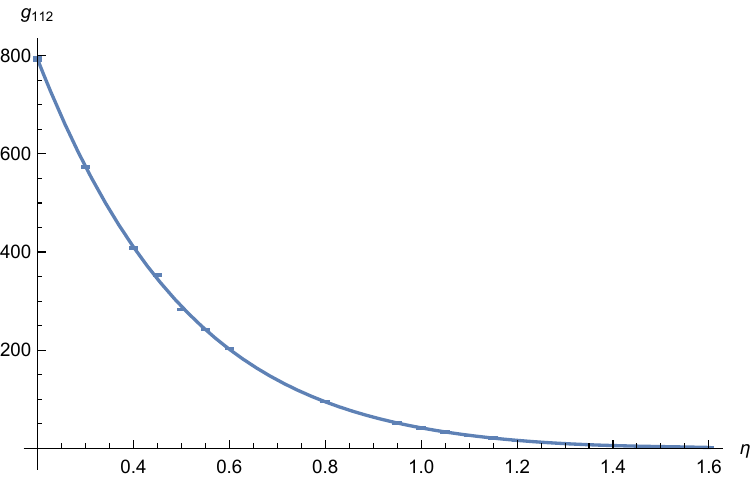}
	\caption{The 3-point couplings $g_{111}$ and $g_{112}$ as functions of $\eta$, determined numerically from the analytically continued $S(z)$ through method II of section \ref{sec:NAC} and then interpolated, with (very small) error bars produced using (\ref{lemmabound}). }
	\label{fig:coupstrength}
\end{figure}

To assume a certain number of resonance zeros may appear to be a gross (and incorrect) simplification. However, the Lemma of section \ref{sec:NAC} gives a rigorous bound on the error of such a fit to $S(z)$ in the interior of the disc, provided that its errors on elastic arc ${\cal I}$, or rather the subset ${\cal I}'$, are known. That is to say, if we could ignore the error in determining the elastic scattering phase and the stable particle masses from TFFSA, we would have a rigorous bound on the analytically continued $S(z)$ despite the presumption on the resonance zeros. If the error bound in the interior of the $z$-disc ends up small, then our assumptions are justified a posteriori. 
On the other hand, if we made incorrect assumptions on the locations of resonance zeros, the error bounds would necessarily be large in the relevant domain, thereby indicating that the fit for $S(z)$ is problematic.

From the residues of $S(z)$ at its poles, we can extract the on-shell 3-point coupling $g_j$ between two lightest particles and the $j$-th lightest particle via \cite{Paulos:2016but}
\ie
g_{11j}^2 
= \frac{4 m_j^3 \left(4m_1^2-m_j^2\right)^{3\over 2}}{m_1^4(m_j^2-2m_1^2)}\lim_{z\to z_j} {z-z_j\over 1-z_jz}S(z),
\fe
where $z_j$ is the pole position corresponding to the mass $m_j$ of the exchanged particle in either $s$ or $t$-channel. The error bound on our numerical approximation to $S(z)$ in the disc allows us to bound the error of the residue. 
The results for $g_{111}$ and $g_{112}$ at positive $\eta$ are shown in Figure \ref{fig:coupstrength} (note that the second stable particle only exists up to a finite value of $\eta$).

\begin{figure}[h!]
	\centering
	\includegraphics[width=.45\linewidth]{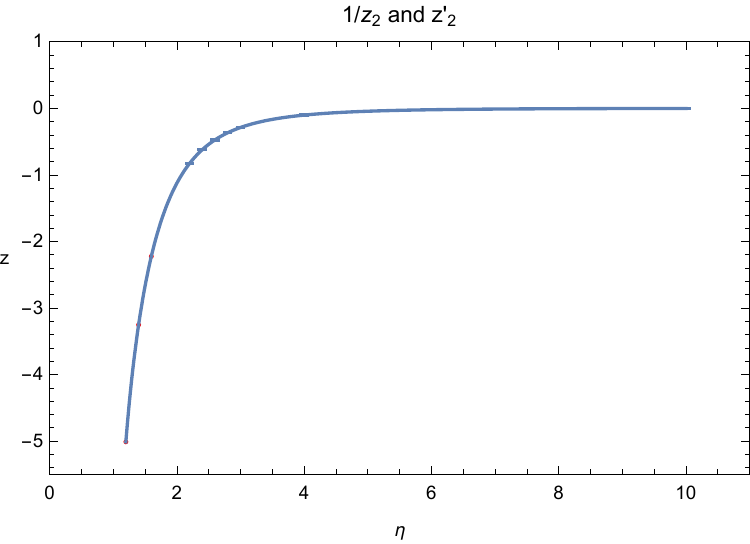}~~~~~
	\includegraphics[width=.45\linewidth]{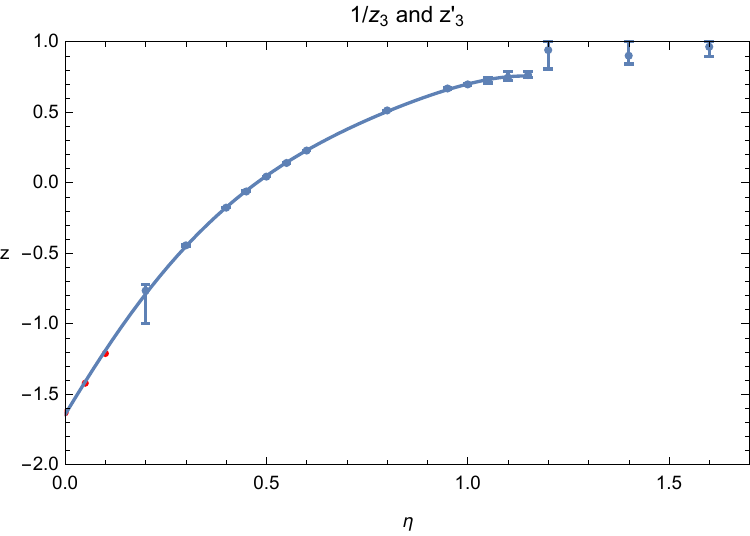}
	\caption{Left: location of the resonance zero $z_2'$ (for $\eta\gtrapprox 2$, blue dots) or the pole $1/z_2$ corresponding to the second lightest particle (for $\eta\lessapprox 2$, red dots, {values for the several lowest $\eta$'s are omitted for clarity}) as a function of $\eta$. Right: location of the resonance zero $z_3'$ (for $\eta\gtrapprox 0.1$, blue dots) or the pole $1/z_3$ corresponding to the third lightest particle (for $\eta\lessapprox 0.1$, red dots) as a function of $\eta$. The pole positions $z_2, z_3$ are determined from the energy of 1-particle states via TFFSA.
	The zeros $z_2', z_3'$ are determined from the numerical analytic continuation of $S(z)$, with error bars inferred from (\ref{lemmabound}).
	}\label{fig:poleEvolution}
\end{figure}

As $\eta$ increases past certain positive values, the second lightest particle (of mass $m_2$) and the third lightest particle (of mass $m_3$) turn into resonances, as will be discussed in section \ref{sec:zeroevol}. These resonances will appear as simple zeros of $S(z)$ on the real axis at $z_2'$ and $z_3'$ respectively. The numerical determination of $z_2'$ and $z_3'$, together with error bars obtained from (\ref{lemmabound}), are shown in Figures \ref{fig:poleEvolution}. Note that the graphs for $z_2'$ and $z_3'$ as functions of $\eta$ smoothly join those of $1/z_2$ and $1/z_3$ ($z_j$ being the pole associated with the $j$-th stable particle), as expected from the poles moving into the second sheet.

The fourth lightest particle (of mass $m_4$), on the other hand, gives rise to a pair of resonance zeros $z_4'$ and $z_4'^*$ at complex values near the boundary of the $z$-disc. Our numerical results for $z_4'$ over a range of $\eta$ is shown in the LHS of Figure \ref{fig:resonanceEvolution}.

\begin{figure}[h!]
	\centering
	\includegraphics[width=.24\linewidth]{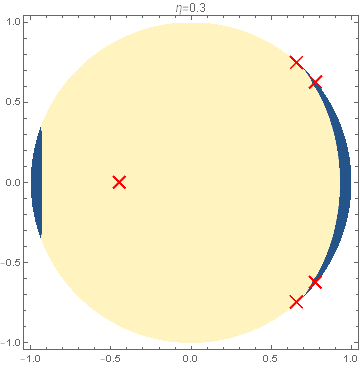}~
	\includegraphics[width=.24\linewidth]{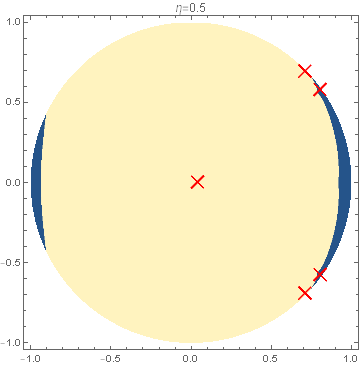}~
	\includegraphics[width=.24\linewidth]{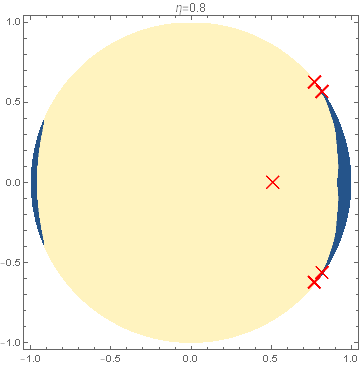}~
	\includegraphics[width=.24\linewidth]{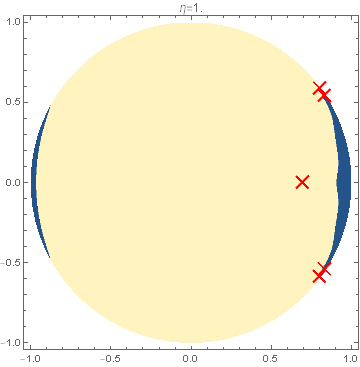}
	\centering
	\includegraphics[width=.24\linewidth]{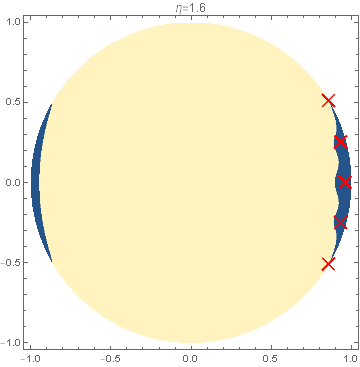}~
	\includegraphics[width=.24\linewidth]{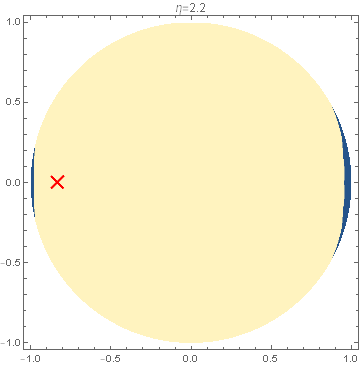}~
	\includegraphics[width=.24\linewidth]{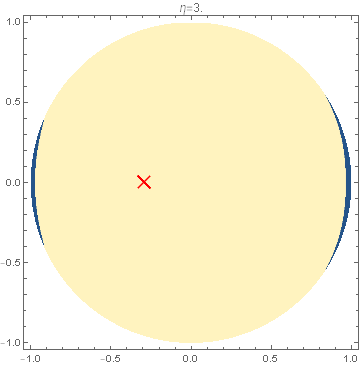}~
	\includegraphics[width=.24\linewidth]{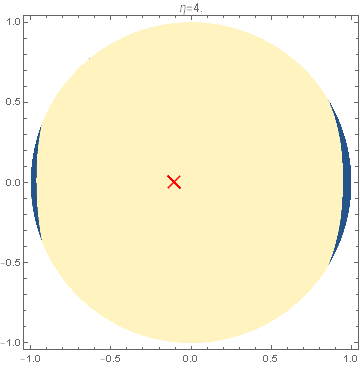}
	\caption{The exclusion plots of unknown resonances on the $z$-disc, for a set of $\eta$ values ranging from $0.2$ to 4. The possibility of zeros of $S(z)$ is eliminated in the yellow region. The zeros assumed for the fit are marked by red crosses.}
	\label{fig:zerExcl}
\end{figure}

\begin{figure}[h!]
	\centering
	\includegraphics[width=.35\linewidth]{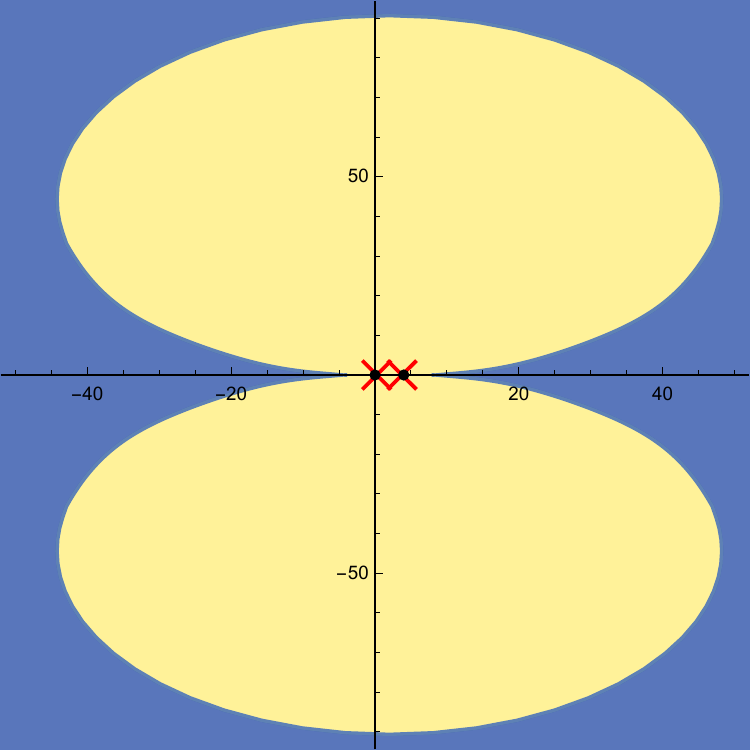}
	\caption{A view of the exclusion plot on the $s$-plane (in units of $m_1^2$), in the case $\eta=3$. The known zeros of $S(s)$ are marked with red crosses (which are very close to the branch points $s=0$ and $4m_1^2$, marked by black dots). The possibility of unknown resonances is eliminated in the yellow region.}
	\label{fig:zerExclsplane}
\end{figure}

While there may well be other resonance zeros that are not taken into account by our approximation to $S(z)$, the error bound (\ref{lemmabound}) allow us to exclude their presence in much of the $z$-disc. Such exclusion plots are shown in Figure \ref{fig:zerExcl}, where unknown resonances are excluded in the yellow region but are possible in the blue ``non-exclusion" regions. For a view of the analogous exclusion plot on the $s$-plane, see Figure \ref{fig:zerExclsplane}.
Note that our error bound is loose near $z=-1$ and near $z=1$, giving rise to non-exclusion regions near both. The non-exclusion region near $z=-1$ is presumably an artifact of our approach, as Luscher's method does not allow us to access the elastic scattering at energies very close to $2m_1$. It is very unlikely that there would be unknown low energy resonances near $z=-1$. On the other hand, there may well be unknown resonances in the non-exclusion region near $z=1$ (high energy limit).

For $0.2<\eta\le 1.6$, it turns out that including an additional pair of complex zeros $z_x'$ and $(z_x')^*$, corresponding to a narrow resonance with energy above the inelastic threshold, improves the fit for $S(z)$ and reduces the error bounds. While our fit suggests the existence of such a resonance, the latter cannot be justified rigorously through our error bound. In particular, the non-exclusion region where we cannot rule out the existence of extra resonance zeros has a connected domain that contains both $z_x'$ and the inelastic arc ${\cal J}$, as shown in the RHS of Figure \ref{fig:resonanceEvolution}. In contrast, the other known resonances (including $z_4'$) lie in their own ``islands" of non-exclusion regions (these islands are too small to be visible in Fig \ref{fig:zerExcl}).  A comparison of the error bounds for $z S(z)$ at real $z$ with or without including $z_x'$ in the fit is shown in Figure \ref{fig:cmpwwoutz_x}.

\begin{figure}[h!]
	\centering
	\subfloat[~~~]{\includegraphics[width=.24\linewidth]{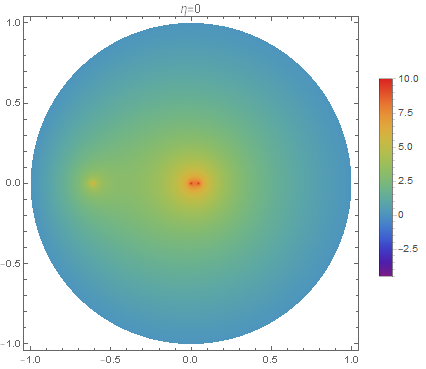}\label{subfig:diska}}~
	\subfloat[~~~]{\includegraphics[width=.24\linewidth]{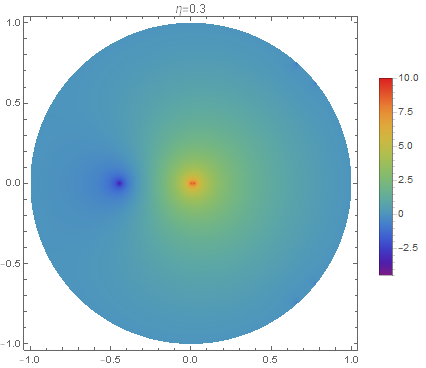}\label{subfig:diskb}}~
	\subfloat[~~~]{\includegraphics[width=.24\linewidth]{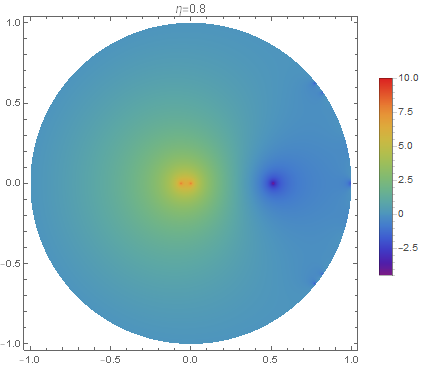}\label{subfig:diskc}}~
	\subfloat[~~~]{\includegraphics[width=.24\linewidth]{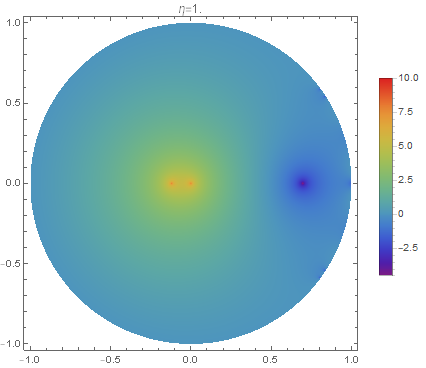}\label{subfig:diskd}}\\
	\centering
	\subfloat[~~~]{\includegraphics[width=.24\linewidth]{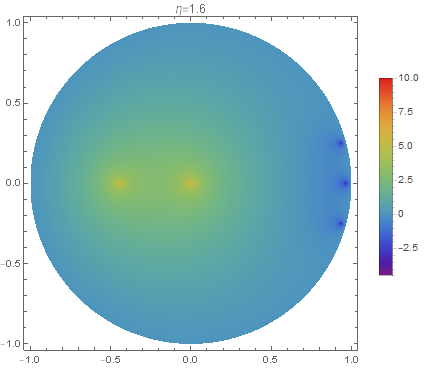}\label{subfig:diske}}~
	\subfloat[~~~]{\includegraphics[width=.24\linewidth]{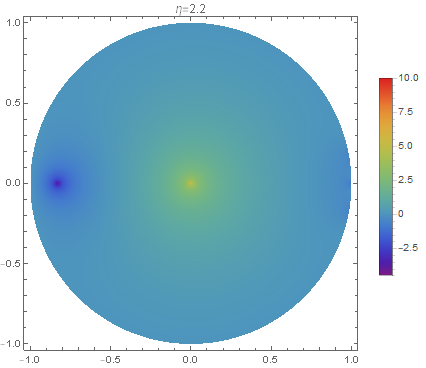}\label{subfig:diskf}}~
	\subfloat[~~~]{\includegraphics[width=.24\linewidth]{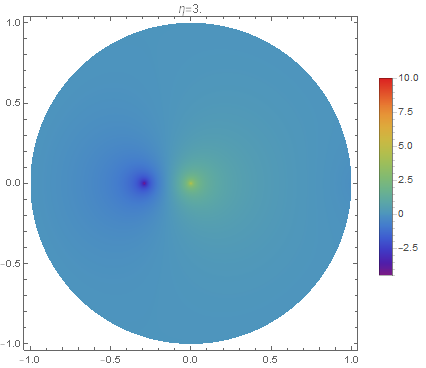}\label{subfig:diskg}}~
	\subfloat[~~~]{\includegraphics[width=.24\linewidth]{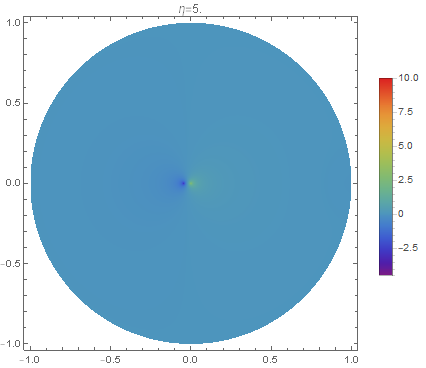}\label{subfig:diskh}}
	\caption{The approximation for $\log|S(z)|$ on the $z$-disc, for various $\eta$ values between 0 and 5, obtained using method II of section \ref{sec:NAC} (error bounds not shown). At each value of $\eta$, we have assumed a certain number of CDD zeros in the fitting, similarly to Figure \ref{fig:zerExcl}. For $\eta>0$, one zero on the real $z$-axis is assumed (this is presumably the resonance $z_3'$ for $0<\eta\le1.6$ and $z_2'$ for $\eta>2$). For $\eta\le1.8$, we include two pairs of complex CDD zeros and fit their positions. As shown in more detail in Figure \ref{fig:resonanceEvolution}, the existence of one of the complex zeros $z_4'$ is justified rigorously by the error bound, and is interpreted as the resonance coming from the particle of mass $m_4$ at the $E_8$ point. The interpretation of the other complex zero $z_x'$ is unclear.
Note that the $z_4'$ and $z_x'$ resonances are not visible in most of the plots due to their proximity to the boundary, except in the $\eta=1.6$ case where $z_x'$ is clearly visible.}
	\label{fig:DscPlots}
\end{figure}

Finally, our numerical results for $\log|S(z)|$ on the entire unit disc and $z S(z)$ on the real interval $(-1,1)$ are shown in Figure \ref{fig:DscPlots} and Figure \ref{fig:eta_RealLine} respectively, with the error bounds indicated in the latter.

\subsection{Evolution of poles and zeros}
\label{sec:zeroevol}

We now describe a simple scenario for the evolution of the poles and zeros of $S(z)$ as functions of $\eta$, which was proposed in \cite{Zamolodchikov:2013ama} and is corroborated by our results. We focus on positive real values of $\eta$, interpolating between the $E_8$ affine Toda theory at $\eta=0$ and the free fermion point in the high temperature limit $\eta\to +\infty$.

Starting at $\eta=0$, where the IFT is integable (Figure \ref{subfig:diska}), $S(z)$ is known to be given exactly by the CDD factors associated with three poles. Apart from the self-coupling pole at $z=0$, there is a CDD pole at $z_2\approx 0.0467965$ corresponding to the second lightest particle with $m_2 = 2\cos {\pi\over 5} m_1$, and another CDD pole at $z_3\approx -0.612751$ corresponding to the third lightest particle with $m_3 = 2\cos{\pi\over 30} m_1$. There is no particle production and no resonance in this case. The infinite energy limit of the S-matrix element is $S(z=1)=1$.

\begin{figure}[h!]
	\centering
	\includegraphics[width=.45\linewidth]{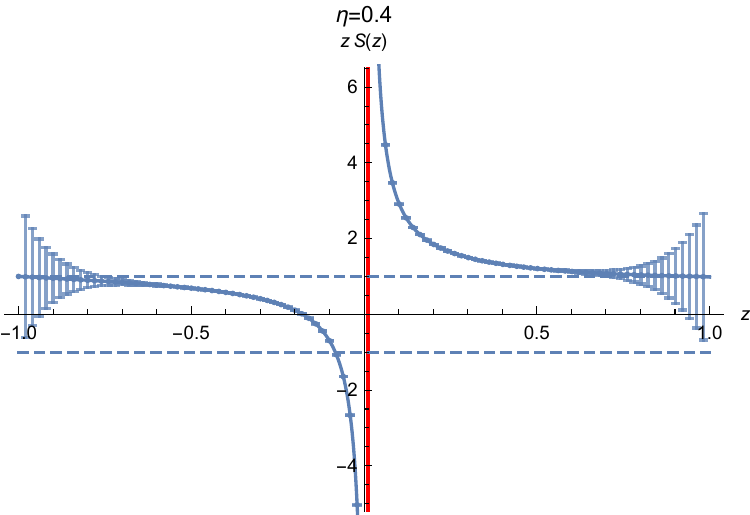}~~~~~
	\includegraphics[width=.45\linewidth]{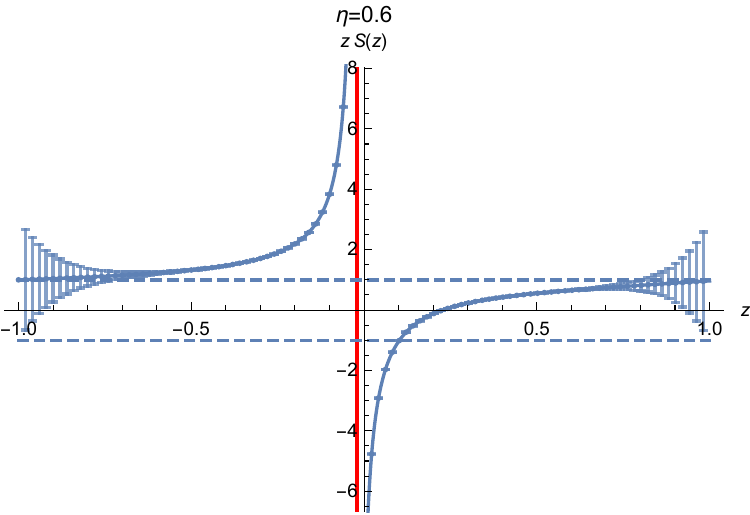}
	\centering
	\includegraphics[width=.45\linewidth]{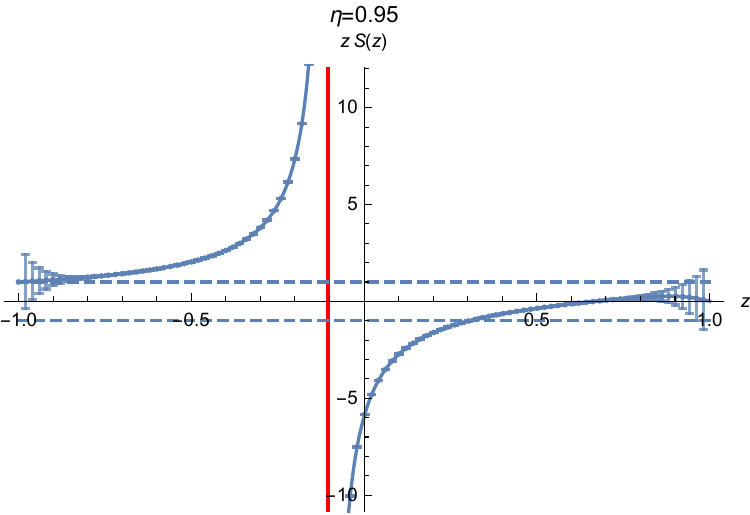}~~~~~
	\includegraphics[width=.45\linewidth]{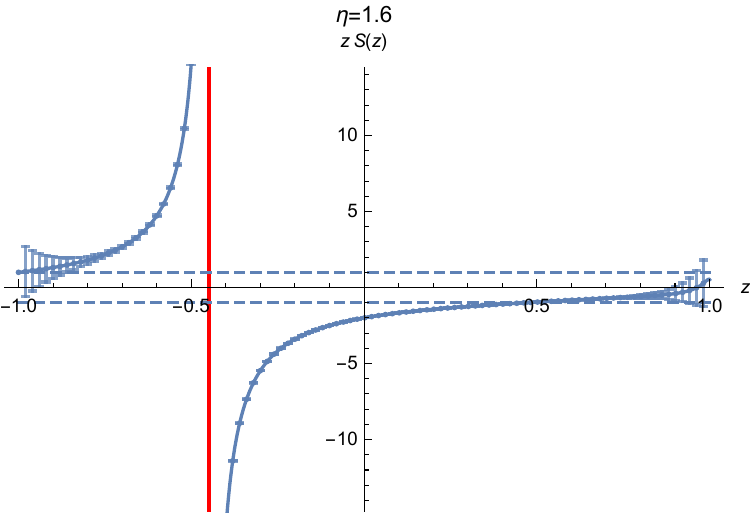}
	\centering
	\includegraphics[width=.45\linewidth]{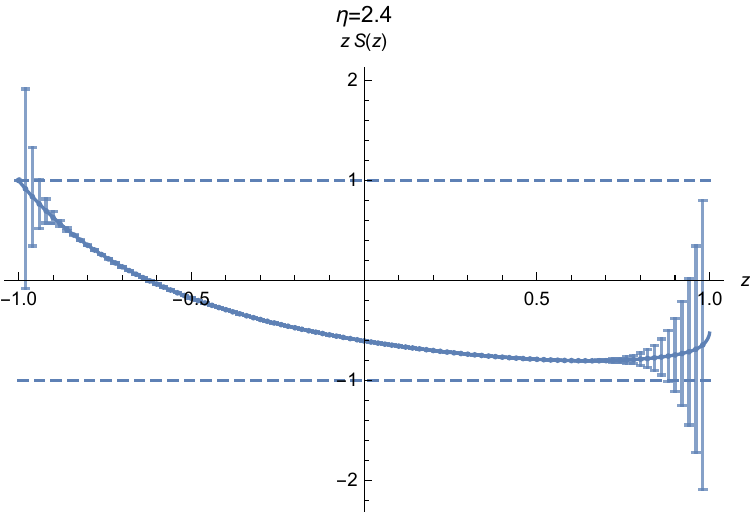}~~~~~
	\includegraphics[width=.45\linewidth]{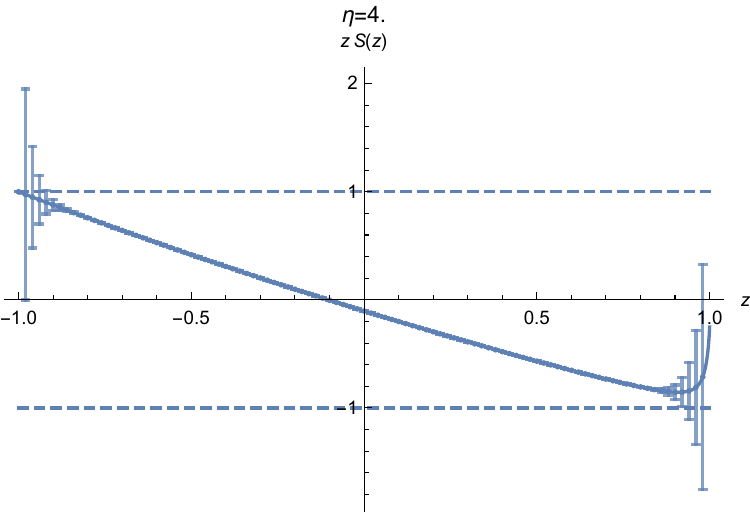}
	\caption{Numerical results for $zS(z)$ and error bounds for real $z\in (-1, 1)$ at various $\eta$. 	}
	\label{fig:eta_RealLine}
\end{figure}

A pole of $S(z)$ can only leave the disc by meeting a mirror zero on the second sheet at $z=-1$ (see Figures \ref{fig:DscPlots}, \ref{fig:eta_RealLine}), as the corresponding particle becomes unstable.
Note that $S(z=-1)=-1$ is maintained throughout. When a pole leaves the first sheet, its mirror zero enters the disc as a resonance zero. In particular, the third lightest particle (of mass $m_3$) decays at $\eta\approx 0.144$, where the pole $z_3$ moves out of the disc and the zero $z_3'$ enters. As $\eta$ increases, $z_3'$ moves along the real $z$-axis with increasing $\eta$ (Figures \ref{subfig:diskb}-\ref{subfig:diske}) toward $z=1$.

We do not know the fate of $z_3'$ at large $\eta$. There are three possible scenarios. One possibility is that $z_3'$ reaches $z=1$ at a finite value of $\eta$ (around $\approx 1.8$ by our crude estimate) and then disappears from the disc. Another possibility is that $z_3'$ ``stays" at $z=1$ rather than ``exits" the disc, and that $S(z)$ vanishes at $z=1$ for $\eta$ greater than some finite value (the latter is suggested by \cite{Zamolodchikov:2011wd}). The third possibility is that $z_3'$ meets another zero of $S(z)$ on the real $z$-axis, and become a pair of complex resonances which then move toward the inelastic arc on the boundary, as $\eta$ increases. This scenario is quite plausible as it is similar to the expected behavior of complex resonances at negative $\eta$ (discussed at the end of this section).

\begin{figure}[h!]
	\includegraphics[width=0.44\linewidth]{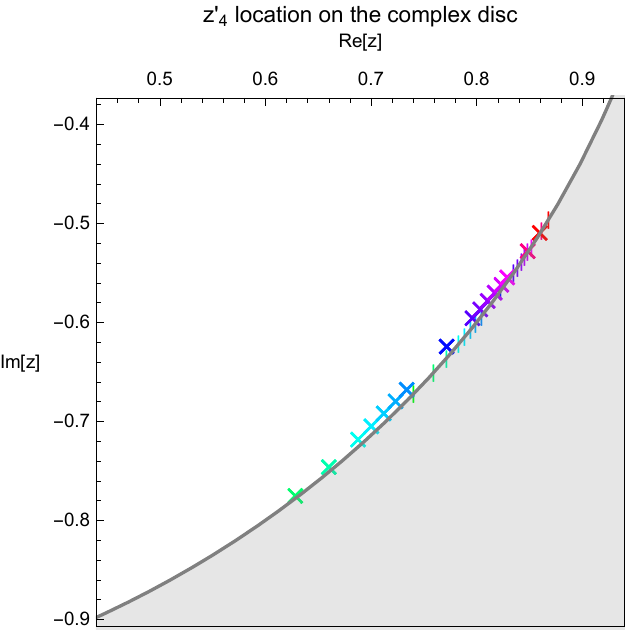}~~
	\includegraphics[width=0.555\linewidth]{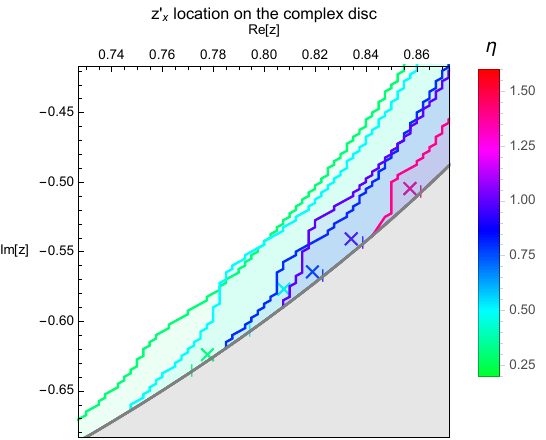}
	\caption{The location of the complex resonance zeros (colored crosses) $z_4'$ (LHS) and $z_x'$ (RHS) for $\eta$ ranging between 0 and $1.6$ (indicated by the color). The white region is part of the unit $z$-disc. Colored vertical stubs indicate the locations of the inelastic threshold. For $z_x'$ we also present the relevant non-exclusion regions in the same colors (these are the same as the blue regions of figure \ref{fig:zerExcl}, zoomed-in near $z_x'$). In particular, $z_x'$ lies in the same connected component of the non-exclusion region as the inelastic arc, and consequently its existence cannot be rigorously justified.}
	\label{fig:resonanceEvolution}.
\end{figure}

The second lightest particle (of mass $m_2$) becomes unstable at $\eta\approx 2.078$, where the pole $z_2$ moves out of the disc and the zero $z_2'$ enters from $z=-1$. As $\eta$ increases further, $z_2'$ moves to the right along the real $z$ axis (Figures \ref{subfig:diskf}-\ref{subfig:diskh}), and approaches $z=0$ as $\eta\to \infty$. In the latter limit, $z_2'$ annihilates the self-coupling pole at $z=0$, as the lightest particle becomes a free fermion.

Simultaneously, as $\eta$ becomes nonzero, the 5 heavier particles of the $E_8$ theory immediately decay and give rise to complex resonances \cite{Zamolodchikov:2013ama,Delfino:2005bh}. In particular, the mass $m_4$ of the fourth lightest particle is below the inelastic threshold at $\eta=0$. For a range of positive $\eta$, the corresponding resonance zero $z_4'$ (and its complex conjugate $z_4'^*$) is close to the boundary of the $z$-disc, and is visible as a jump by $2\pi$ in the elastic scattering phase (see Figure \ref{fig:eta_infLvlFit} for $\eta=0.3$ and $0.8$). We can take into account such a pair of CDD zeros in the ansatz for fitting $S(z)$ (via method II of section \ref{sec:NAC}), which then allows us to determine the position of $z_4'$ up to a small known bound on its error.

\begin{figure}[h!]
	\centering
	\includegraphics[width=0.45\linewidth]{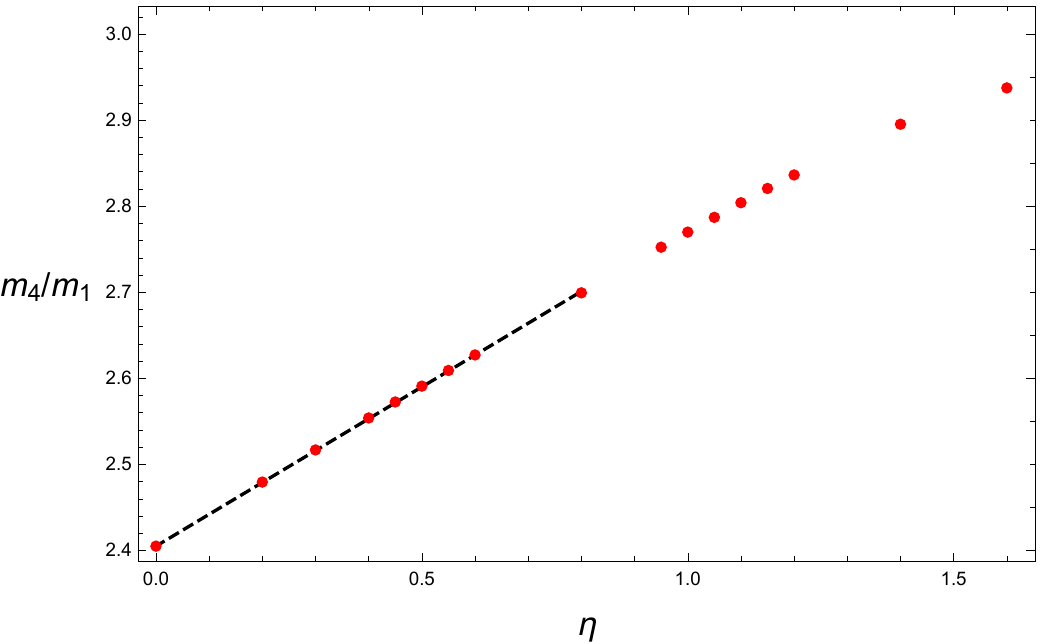}~~~~~	\includegraphics[width=0.54\linewidth]{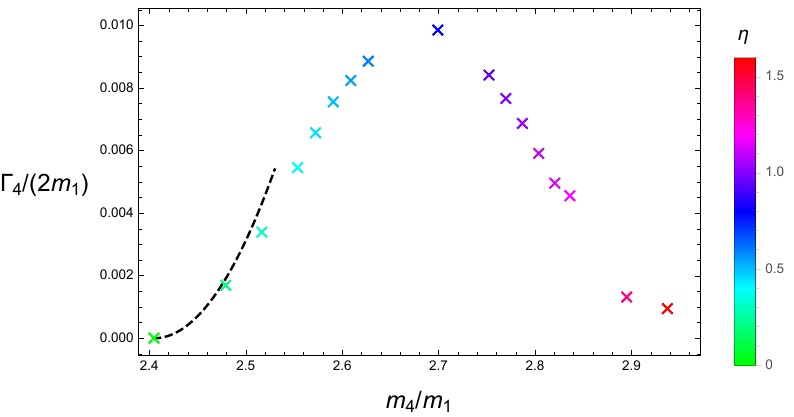}
	\caption{Left: $m_4/m_1$ as a function of $\eta$, where $m_4$ is the real part of the energy of the $z_4'$ resonance. The red dots represents ${\rm Re} \sqrt{s}/m_1$, where $s$ is the locations of the complex resonance $z_4'$ on the $s$-plane. The black dashed line is the first order result of form factor perturbation theory around the $E_8$ point $\eta=0$. Right: $\Gamma_4/(2m_1)$ as a function of $m_4/m_1$, where the width $\Gamma_4$ is twice the imaginary part of the energy of the $z_4'$ resonance. The crosses represent the locations of $\sqrt{s}/m_1$ on the complex plane for the $z_4'$ resonance (equivalent to those in the $z$-disc plot in left of Figure \ref{fig:resonanceEvolution}). The black dashed curve represents the leading order perturbation theory result \cite{Delfino:1996xp,Pozsgay:2006wb} (where $\Gamma_4/(2m_1)$ is computed to order $\eta^2$, and $m_4/m_1$ to order $\eta$, near $\eta=0$).} \label{fig:cmpzz4WithPert}.
\end{figure}

The evolution of $z_4'$ with $\eta$ between 0 and $1.6$ is portrayed in the LHS of Figure \ref{fig:resonanceEvolution}. It begins at the point corresponding to $m_4=4\cos{\pi\over 5}\cos{7\pi\over 30}$, and stays close to the elastic arc for $\eta$ up to $\approx 2$. Note that the inelastic threshold $m_1+m_2$ moves with $\eta$ until $m_2$ reaches $2m_1$ and the second lightest particle decays. In Figure \ref{fig:cmpzz4WithPert}, we show the real and imaginary part of the energy of the $z_4'$ resonance for positive $\eta$, with a comparison to results from leading order form factor perturbation theory around $\eta=0$ \cite{Delfino:1996xp,Pozsgay:2006wb}.\footnote{We thank G. Mussardo and G. Tak\'acs for bringing the results of \cite{Delfino:1996xp,Pozsgay:2006wb} to our attention.}

For $\eta\leq 1.8$, we find reasonably good fitting for $S(z)$ with the additional complex resonance zero $z_x'$ included, but with large uncertainty in the location of $z_x'$ (RHS of Figure \ref{fig:resonanceEvolution}). We are unable to determine reliably the location or the existence of $z_x'$ or any higher energy resonances, e.g. those coming from the more massive particles at the $\eta=0$ point, corresponding to pairs of complex zeros, or other possible real zeros near $z=1$.

\begin{figure}[h!]
	\centering
	\includegraphics[width=.325
	\linewidth]{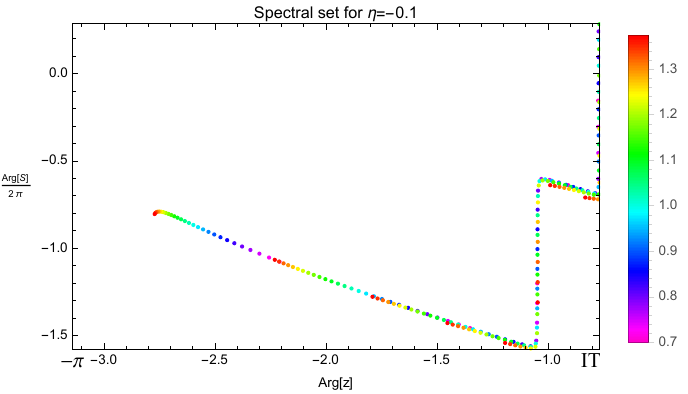}
	\includegraphics[width=.325
	\linewidth]{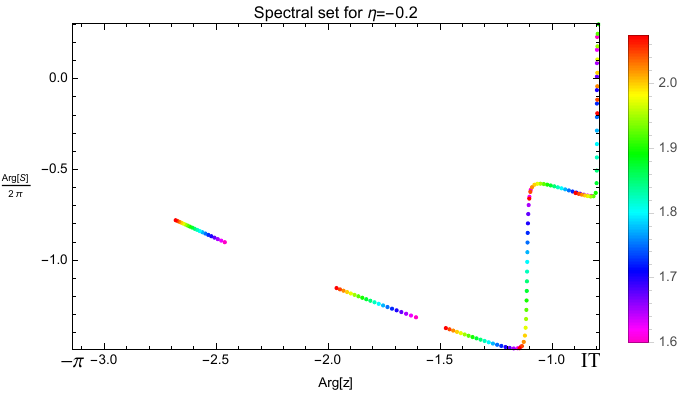}~
	\includegraphics[width=.325
	\linewidth]{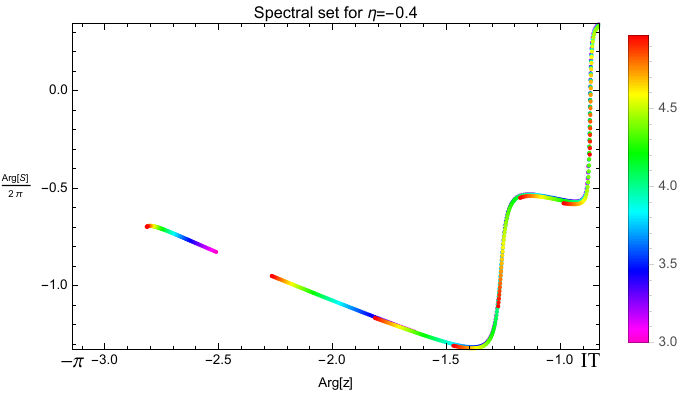}
	\centering
	\includegraphics[width=.325
	\linewidth]{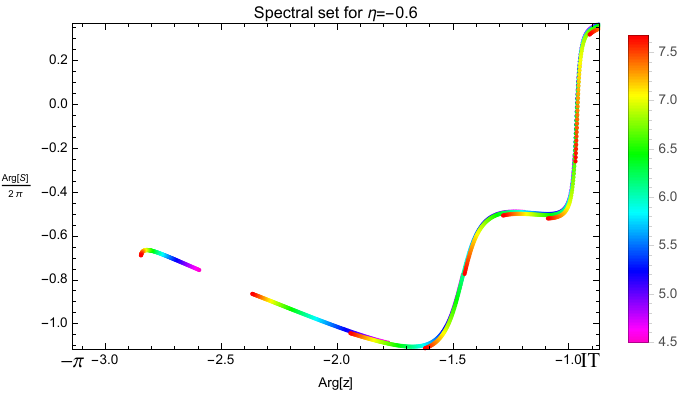}~
	\includegraphics[width=.325
	\linewidth]{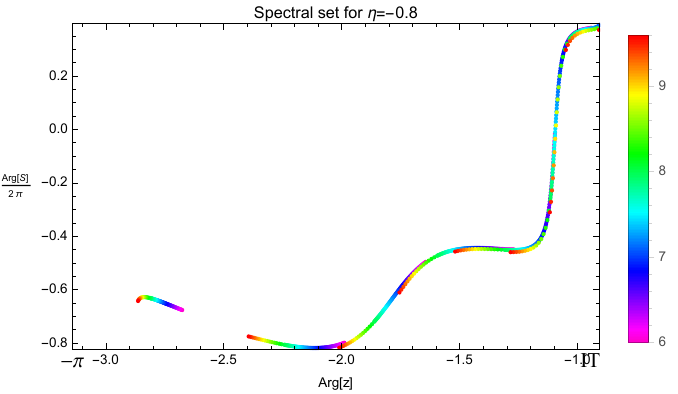}
	\includegraphics[width=.325
	\linewidth]{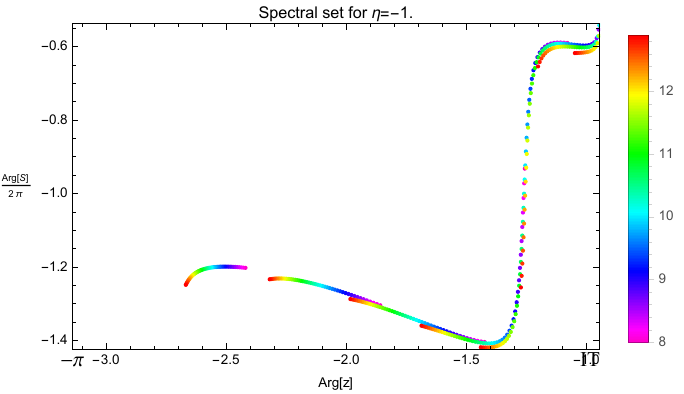}
	\caption{Spectral sets for the elastic scattering phase, computed from TFFSA with truncation level $L=21$ over a range of cylinder radii (indicated by color), for some $\eta$ values between $-1$ and $0$.}
	\label{fig:negetareso}
\end{figure}

As for the ``low temperature" $\eta<0$ regime, where more stable particles (``mesons") are present in the spectrum of IFT, some preliminary results on the elastic scattering phase 
are presented in Figure \ref{fig:negetareso}. Here we see evidence for the way in which the fourth stable particle enters the spectrum, suggested in \cite{Fonseca:2006au}. Namely, as $\eta$ is decreased below 0, the $m_4$ resonance becomes wider and moves to lower energies, until its corresponding pair of complex zeros in the $z$-disc meet on the real $z$-axis. As $\eta$ is decreased further, one of the zeros moves in the positive-$z$ direction, remaining a resonance, whereas the other zero moves toward $z=-1$. The latter eventually exits the disc, as its mirror pole enters at $z=-1$ from the second sheet, giving rise to the stable $m_4$ particle. The $m_5$ resonance is visible in Figure \ref{fig:negetareso} as well. Detailed corroboration and further analysis of the low temperature regime are of interest and are left for future work.

\subsection{Comparison to perturbation theory}

In the elastic scattering regime, that is, for $|\theta|<\theta_0$ where $\theta_0$ is the rapidity at the inelastic threshold, the deviation of the scattering phase from that of the CDD factor is related to the function $f(\theta)$ in (\ref{sgenstr}) via
\ie\label{deltphisa}
\Delta\varphi(\theta)\equiv {\varphi(\theta) - \varphi_{\rm CDD}(\theta)} = \int_{\theta_0}^\infty {d\theta'\over 2\pi} \left[{1\over \sinh(\theta-\theta')} + {1\over \sinh(\theta+\theta')} \right] \log f(\theta').
\fe
Note that in this regime (\ref{deltphisa}) is positive and monotonically increasing.
$f(\theta)$ is further related to the total inelastic scattering probability $\sigma_{\rm tot}(\theta)$ via
\ie\label{fsigma}
f(\theta) = 1 - \sigma_{\rm tot}(\theta).
\fe

\begin{figure}[h!]
	\centering
	\includegraphics[width=.45
	\linewidth]{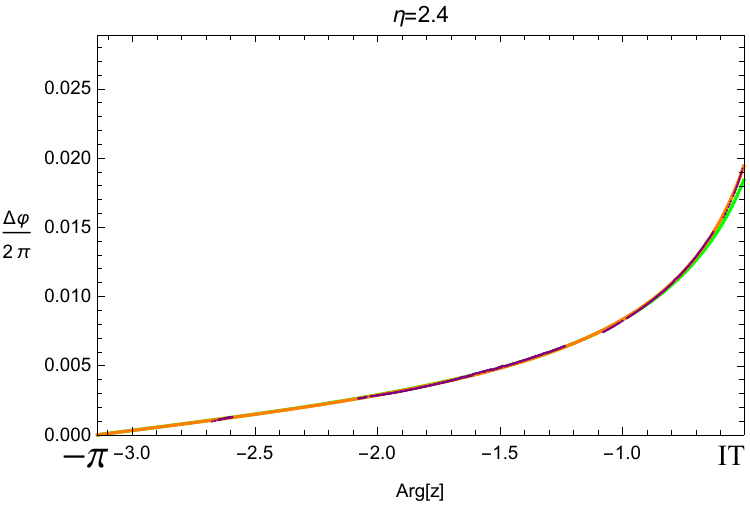}~~~~~
	\includegraphics[width=.45
	\linewidth]{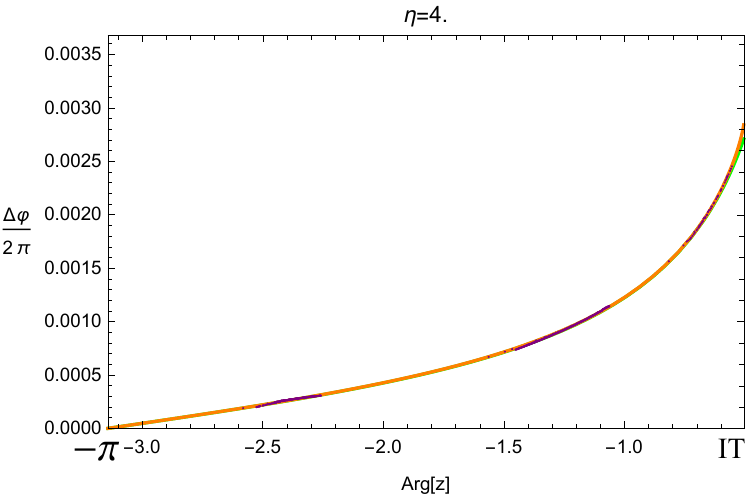}
	\caption{The non-CDD part of the elastic scattering phase $\Delta\varphi$ as a function of $\arg(z)$. The orange curve is obtained by fitting the phase of (\ref{phasfit}). The purple curve segments are obtained by subtracting the fitted {\it known} CDD factors from the elastic scattering phase (spectral sets (\ref{eq:specstets}) computed from TFFSA over suitable ranges of cylinder radii, extrapolated to infinite truncation level). In the $\eta=2.4$ case, the leading perturbative result for the non-CDD part, based on the $2\to 3$ probability (\ref{sigmatt}) is shown in green curve. In the $\eta=4$ case, the perturbative result (green curve) is indistinguishable from the orange curve.}
	\label{fig:eta10cmppert}
\end{figure}

For weak magnetic field $h$ in the high temperature regime, namely $\eta\gg 1$, there is only one stable particle, and $\sigma_{\rm tot}(\theta)$ is dominated by the $2\to 3$ scattering probability $\sigma_{2\to 3}(\theta)$. The latter is computed to order $h^2$ in \cite{Zamolodchikov:2011wd}, with the result (written in $m_f=1$ units)
\ie\label{sigmatt}
& \sigma_{2\to 3}(\theta) = h^2 K(\theta) + {\cal O}(h^4),
\\
& K(\theta) = {4\cdot 2^{1\over 6} e^{-{1\over 4}} A_G^3\over \pi} {(E+2)^{5\over 2}\over (E-2)^{3\over 2}} {(2E-1)^4 (E-3)^3\over (E+1)(E-1)^{5\over 2}(E+3)^{3\over 2} E^3} \int_{-1}^1 dt \left( {1-\mu t^2\over 1- \nu t^2} \right)^2 {\sqrt{1-t^2}\over (1-\lambda t^2)^{3\over 2}},
\fe
where $A_G$ is Glaisher's constant, $E\equiv 2\cosh {\theta\over 2}$, and $\mu, \nu, \lambda$ are defined as
\ie
\mu \equiv {(E-2)(2E+1)^2\over (E+2)(2E-1)^2}\lambda, ~~~\nu \equiv {E+2\over E-2}\lambda,~~~\lambda\equiv {(E+1)(E-3)^3\over (E-1)(E+3)^3}.
\fe
The leading perturbative contribution to $\Delta\phi(\theta)$ is 
\ie\label{pertdevia}
- \eta^{-{15\over 4}}\int_{\theta_0}^\infty {d\theta'\over 2\pi} \left[ {1\over \sinh(\theta-\theta')}+{1\over \sinh(\theta+\theta')} \right]K(\theta').
\fe

A comparison between our numerical fit for the (non-perturbative) S-matrix element $S(z)$ and the perturbative result based on (\ref{sigmatt}), (\ref{pertdevia}) is shown in Figure \ref{fig:eta10cmppert}. The agreement is close for $\eta=2.4$ and nearly perfect for $\eta=4$.

\subsection{Constraining the inelastic cross section}
\label{sec:inelastichighenergy}

In applying method II to determine the analytic continuation of $S(z)$ from the elastic arc ${\cal I}$ to the interior of the $z$-disc, we have fitted the function $f(\theta)$ appearing in (\ref{sgenstr}) or (\ref{soverscdd}). Naively, this could give a numerical estimate of the inelastic scattering probability via (\ref{fsigma}). However, the bound (\ref{lemmabound}) becomes loose when $z$ approaches the inelastic arc ${\cal J}$ on the boundary of the disc, and we cannot rigorously bound the error on the fit for $f(\theta)$. Our inability to exclude narrow resonances close to the inelastic arc $\mathcal J$ is the other side of the same coin, as their effects on the elastic phase can hardly be distinguished numerically from (\ref{deltphisa}) due to inelastic processes.\footnote{For instance, had we missed a CDD zero $z_*'$ (and its complex conjugate) close to the inelastic arc ${\cal J}$, with $z_*' = {\sinh (\theta_* + i \delta_*) - {\sqrt{3}\over 2} i\over \sinh (\theta_* + i \delta_*) + {\sqrt{3}\over 2} i}$ for real $\theta_*>\theta_0$ and small positive $\delta_*$, the corresponding CDD factor
\ie
{\sinh(\theta_* + i \delta_*) - \sinh\theta\over \sinh(\theta_* + i \delta_*) + \sinh\theta} \cdot {\sinh(-\theta_* + i \delta_*) - \sinh\theta\over \sinh(-\theta_* + i \delta_*) + \sinh\theta} = e^{ - 2 i \delta_* \left[ {1\over \sinh(\theta-\theta_*)} + {1\over \sinh(\theta+\theta_*)} \right] + {\cal O}(\delta_*^3)}
\fe
would imitate a contribution to $-\log f(\theta)$ by $4\pi \delta_* \delta(\theta-\theta_*) + {\cal O}(\delta_*^3)$, resulting in an extra ``resonance peak" in the numerical fit for $-\log f(\theta)$.}

\begin{figure}[h!]
	\centering
	\includegraphics[width=.45
	\linewidth]{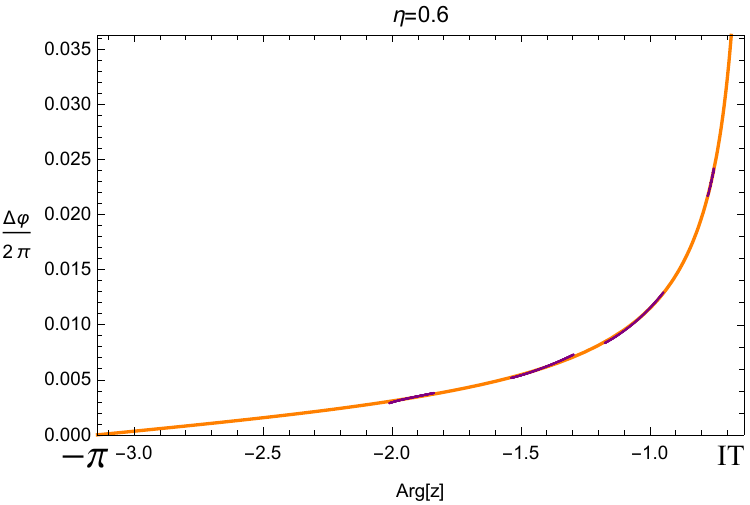}~~~~~
	\includegraphics[width=.45
	\linewidth]{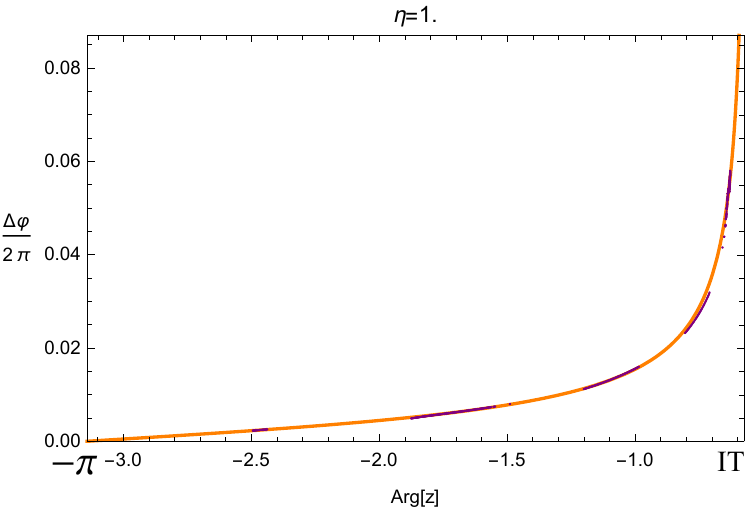}
	\caption{Upper bound on $\Delta\varphi$ (\ref{intetransf}) as a function of $\phi\equiv\arg(z)$, at smaller $\eta$ values where more than one stable particle exists. The bound is obtained by fitting the non-CDD part of the elastic phase, similarly to that of Figure \ref{fig:eta10cmppert}, taking into account the known resonance zeros including $z_4'$ (but without including $z_x'$). 
	}
	\label{fig:boundsOnArgSf}
\end{figure}

It is nonetheless possible to place reliable constraints on 
the inelastic cross section suitably integrated over energies, as follows. While we have not taken into account all CDD zeros in the numerical fit, missing CDD zeros can only correct the ``non-CDD elastic scattering phase" $\Delta\varphi$ defined in (\ref{deltphisa}) by a negative amount. 
Hence our numerical fit for $\Delta\varphi$ can be thought of as an upper bound on the following integral transform of the inelastic cross section $\sigma_{\rm tot}$ (here parameterized as a function of $\arg(z)$ rather than the rapidity)
\begin{equation}
\label{intetransf}
\Delta\varphi (\phi) = \int\limits_{0}^{\phi_0}\frac{d\phi'}{\pi}  \frac{\sin (\phi)}{\cos (\phi')-\cos (\phi)} \log {1\over 1-\sigma_{\rm tot}(\phi')},
\end{equation}
for $\phi\in (\phi_0, \pi)$ (the elastic arc).

Examples of such upper bounds are shown in Figure \ref{fig:boundsOnArgSf}. Here we have computed the non-CDD part of the elastic scattering phase by fitting $S(z)$ while taking into account the known resonance zeros, including the complex resonance $z_4'$ and $(z_4')^*$, but not including $z_x'$ whose existence has not been firmed established. The result can then be interpreted as a rigorous upper bound on (\ref{intetransf}), subject to numerically negligible uncertainties in the location of the known resonance zeros (as they sit in tiny islands of the non-exclusion region in Figure \ref{fig:zerExcl}), and subject to the error in the elastic data obtained from TFFSA.

\section{Discussion}
\label{discusssec}

Let us recap our method of determining the analytically continued $2\to 2$ S-matrix element $S(z)$ of the lightest particle in IFT. We employed TFFSA to compute the spectrum of IFT on a circle, and used Luscher's method to extract the scattering phase of a pair of particles in the elastic regime. Unitarity and analyticity properties of the S-matrix then constrains $S(z)$ in terms of known CDD poles (from stable particle masses), a priori unknown CDD zeros (resonances), and the inelastic scattering probability $1-f(\theta)$. We make an educated guess on the number and rough location of the CDD zeros, and fit the zeros along with $f(\theta)$ to the elastic scattering data. 

The key lemma presented in section \ref{sec:NAC} justifies our fitting for $S(z)$ a posteriori, as it allows us to rigorously bound the error in the interior of the $z$-disc, assuming that the elastic scattering phase is known accurately. While we may not have account for all CDD zeros in the fitting, we can use the error bar to exclude the presence of unknown zeros in extended regions of the $z$-disc. The results, presented in section \ref{nacsec}, corroborate the simple scenario of the evolution of poles and resonances proposed in \cite{Zamolodchikov:2013ama}. Furthermore, we have determined the location of the real resonance zeros ($z_2'$, $z_3'$) to reasonably good accuracy, as well as the first complex resonance ($z_4'$) provided that the real part of its energy is below the inelastic threshold. Our results for the non-CDD part of the elastic phase are in good agreement with leading order form factor perturbation theory result for $\eta>2$.

Many questions about $S(z)$ remain unanswered. For instance, we do not know the ultimate fate of the $m_4$ resonance as $\eta$ increases above 2, whether it exits the disc (e.g. becoming unbound as a multi-particle state) or continues to exist at large positive $\eta$. It would be interesting to follow the evolution of the four higher resonances as well. Another outstanding question is the behavior of the high energy limit $S(z=1)$ as a function of $\eta$. 

To address these questions with the method of this paper would require more accurate data in the elastic regime. The error in TFFSA is due to the finite truncation level. While one can always try to increase the truncation level, this is computationally expensive and qualitative improvements are unlikely. A more promising approach is to work with an RG-improved truncated Hamiltonian, along the lines of \cite{Giokas:2011ix, Hogervorst:2014rta, Rychkov:2014eea, Rychkov:2015vap}. The errors introduced by Luscher's method, on the other hand, are due to finite size effects. These may be corrected by taking into account wrapping interactions.

While we have not investigated the low temperature regime $\eta<0$ in detail, some interesting features of the elastic scattering phase is already seen in Figure \ref{fig:negetareso}. For sufficiently large negative $\eta$, the stable particles are mesons \cite{Fonseca:2006au} and their scattering at high energies may be captured by a sort of parton or flux-string approximation.

So far we have focused exclusively on the $2\to 2$ scattering of the lightest particle. It would be interesting to extend the analysis to other S-matrix elements, and utilize TFFSA for states with 2 or more particles beyond the elastic regime.

Another avenue of investigation is the analytic continuation of IFT to complex $\eta$. For instance, in \cite{Fonseca:2001dc} TFFSA is applied to the non-unitary IFT with $\eta=y e^{4\pi i/15}$ for real $y$, in which case the eigenvalues of the Hamiltonian are either real or come in complex conjugate pairs. It would be interesting to see whether a similar analysis of the S-matrix can be performed for the non-unitary IFTs.

Finally, one could hope to extend the method of this paper to more than two spacetime dimensions, which would require extending the lemma of section \ref{sec:NAC} to partial wave amplitudes that obey elastic unitarity and crossing symmetry.
 

\section*{Acknowledgements} 

We are grateful to Minjae Cho, Sergei Dubovsky, Joao Penedones, Slava Rychkov, Amit Sever, Jonathan Toledo, Balt van Rees, Pedro Vieira, Alexander Zamolodchikov, and Alexander Zhiboedov for discussions. We thank the organizers of Bootstrap 2018, the Caltech high energy theory group, and the workshop Advances in Quantum Field Theory at CERN for their hospitality during the course of this work. XY thanks the organizers of the Azores workshop on S-matrix bootstrap, Strings 2018, Kavli Institute for Theoretical Physics, Simons Center for Geometry and Physics, and Galileo Galilei Institute for Theoretical Physics for hospitality. This work is supported in part by a Simons Investigator Award from the Simons Foundation, by the Simons Collaboration Grant on the Non-Perturbative Bootstrap, and by DOE grant DE-SC00007870. The numerical computations in this work are performed on the Odyssey cluster supported by the FAS Division of Science, Research Computing Group at Harvard University.

\appendix

\section{Bounding errors inside the disc}
\label{app:strongerLemma}

In this  Appendix we prove the lemma asserted in section \ref{sec:NAC}. Let us write
\ie
\delta = \int_{\cal I} {du\over 2\pi i u} \mu(u)  |F(u)|,
\fe
where $\mu(u)$ is some positive measure factor on ${\cal I}$, to be determined later. We can construct the following analytic function $\chi_{z}(w)$ on the unit disc $|w|<1$,
\ie\label{chiformFull}
\chi_{z}(w) = \exp\left[ \oint_C {du\over 2\pi i} {\log h_{z}(u) \over u} {u+w\over u-w} \right],
\fe
where $C$ is the counterclockwise contour along the unit circle $|u|=1$, and $h_{z}(u)$ is the real positive valued function defined by
\ie\label{hformFull}
& h_{z}(u) = \widetilde\delta^{1-{2\pi\over \ell_{z}}} \mu_z(u),~~~ u \in \tau_z^{-1}({\cal I}),
\\
& h_{z}(u) = e^{- (1 - {\ell_z\over 2\pi})^{-1} A_{z}} \widetilde\delta ,~~~ u \in \tau_z^{-1}({\cal J}),
\fe
for a constant $\widetilde\delta>0$. Here we have defined $\ell_z \equiv \tau_z^{-1}({\cal I})$, and
\ie{}
&A_{z} \equiv  \int_{\tau_{z}^{-1}({\cal I})} {du\over2\pi iu} \log\mu_z(u),
~~~~~~  \mu_z(u) \equiv {u\over \tau_z(u)} {\partial\tau_z(u)\over \partial u} \mu(\tau_z(u)).
\fe
Note that $|\chi_z(u)|$ coincides with $h_z(u)$ along $|u|=1$. Furthermore, (\ref{hformFull}) is such that $\log h_z(u)$ averages to zero on the unit circle, and therefore $\chi_z(0)=1$.

Now consider the identity
\ie\label{fcontrepFull}
F(z) = \oint_C {dw\over 2\pi i} {\chi_z(w)\over w} F(\tau_z(w))
\fe
that holds for any $F(z)$ that is analytic and bounded on $|z|<1$. It follows that
\ie\label{ffdiffFull}
|F(z)| &\leq \int_{\tau_z^{-1}({\cal I})} {dw\over 2\pi i} {h_z(w) |F(\tau_z(w))|\over w} + \int_{\tau_z^{-1}({\cal J})} {dw\over 2\pi i} {h_z(w) |F(\tau_z(w))|\over w}
\\
&\leq  \widetilde\delta^{1-{2\pi\over \ell_z}} \delta + B_z (1 - {\ell_z\over 2\pi}) e^{- (1 - {\ell_z\over 2\pi})^{-1} A_{z}} \widetilde\delta,
\fe
where $B_z$ is given by (\ref{bzforms}).
Minimizing the RHS with respect to $\widetilde\delta$, we have
\ie
|F(z)| \leq B_z^{1-{\ell_z\over 2\pi}} e^{-A_z} \left({2\pi\delta \over \ell_z}\right)^{{\ell_z\over 2\pi}}.
\fe
So far, the measure $\mu(u)$ is arbitrary, and we can optimize the bound by choosing a suitable $\mu(u)$, or equivalently $\mu_z(u)$, {\it at each given} $z$. For our purpose it will suffice to minimize $e^{-A_z} (2\pi \delta/\ell_z)^{\ell_z\over 2\pi}$, which amounts to taking
\ie
\mu_z(u) = {2\pi \delta_z \over \ell_z |F(\tau_z(u))|} ,~~~~\delta_z\equiv \int_{\tau_z^{-1}({\cal I})} {dw\over i w } |F(\tau_z(w))|.
\fe
This yields the bound (\ref{lemmabound}).

\section{Numerical implementation}

The numerical implementation of our method involves the following steps (with estimated run times and memory usage on the Harvard research computing cluster indicated).

\begin{enumerate}
	\item \textbf{Form factors at finite radius} - We use Mathematica to evaluate the finite size contributions to the form factor (\ref{eq:FSTrans}). At a given radius, all of the matrix elements are calculated up to a given level. \textit{Runtime: a few hours per radius. Memory: 4GB. Both up to level 22.}
	\item \textbf{TFFSA} - We use Julia to generate the truncated Hamiltonian and diagonalize it numerically. In each instance, a data set is generated for a given coupling, at a given radius and truncation level. \textit{Runtime: a few minutes to several hours. Memory: 1.5 to 100 GB. For levels 16 to 22.}
	\item \textbf{Masses} - The masses of the stable particles are determined using the TFFSA spectra. Each energy eigenvalue corresponding to a 1-particle state is first extrapolated to infinite truncation level, and then extrapolated to infinite radius. \textit{Runtime: a few minutes for each value of $\eta$. Memory: 4GB.}
	\item \textbf{Spectral Sets} - The spectra for all truncation levels are translated into spectral sets, which are then interpolated to smooth curves. We extrapolate the phase curves to infinite level using the fit (\ref{eq:extrapToInf}). The quality of the extrapolation is validated by either one of two tests. The first test demands the quality of the fit to satisfy a discrete Cram\'er-von Mises criterion, i.e. imposing a threshold on the sum of the differences squared between the fit and the finite level truncation data, and demands that the difference between the highest level data and the infinite level extrapolation is small relative to the difference between the highest level and second highest level data. This test is typically used for smaller positive values of $\eta$, and aims to confirm that the level truncation data converge to the extrapolated results. The second test demands that the difference between the extrapolated result and the finite level data is less than some small constant. This is typically used for large $\eta$, where the finite level truncation results have converged already numerically.
	\textit{Runtime: Around 20 hours per value of $\eta$. Memory: 4GB.}
	\item \textbf{Fitting} - The fitting code takes the extrapolated elastic scattering phase data obtained in the previous step, subtracts the CDD pole contributions (determined by the stable particle masses obtained in step 3) and fits to the ansatz (\ref{phasfit}) by minimizing the difference over a dense set of energies (in the elastic regime). This ansatz is based on a finite set of CDD zeros, each of which is presumed to reside in a certain domain, along with a ``non-CDD" exponential factor that involves the function $Q(x)$ approximated by the sum of squares of polynomials of a certain degree. Note that the integration involving $Q(x)$ in (\ref{phasfit}) is carried out analytically. \textit{Runtime: between 10 minutes for fitting $Q(x)$ with $3$ parameters to around 8 hours for $10$ parameters. Memory: 8GB.}
	\item \textbf{Error-bounds and values I}
	We evaluate the fit for $S(z)$ on the elastic arc and on the real $z$-axis (each with between $500$ and $3000$ sampling points, depending on the value of $\eta$). On the real $z$-axis we compute the error bounds using (\ref{lemmabound}), discretizing the relevant integrals with a few thousand sampling points. 
	\textit{Runtime: Around 15 hours for each $\eta$ value. Memory: 12GB.}
	\item \textbf{Error-bounds and values II}
	We evaluate the fit for $S(z)$ and its error bound in the interior of the disk (with $\sim \pi \times 200^2$ sampling points). Due to the number of sampling points we divide this to typically $25$ separate jobs. \textit{Runtime: Around 20 hours per $1/25$ partition. Memory: 24GB.}
	\end{enumerate}


\section{Comparisons with alternative fittings}
\label{app:cmpMethI}

\begin{figure}[h!]
	\centering
	\includegraphics[width=.31\linewidth]{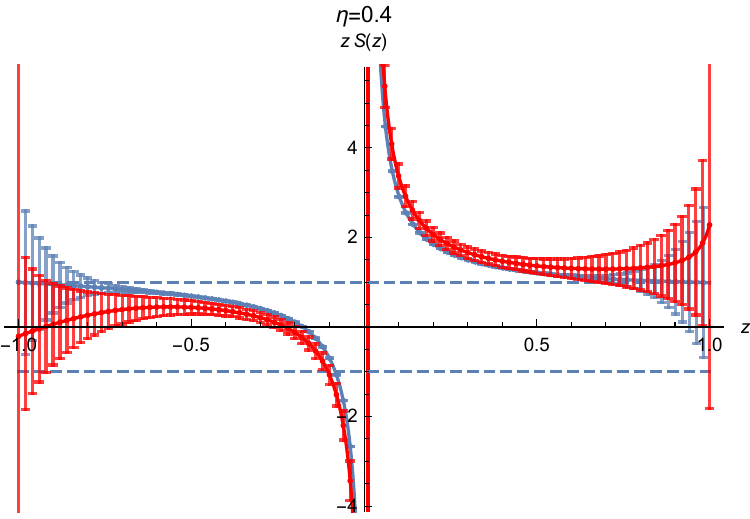}~~~
	\includegraphics[width=.31\linewidth]{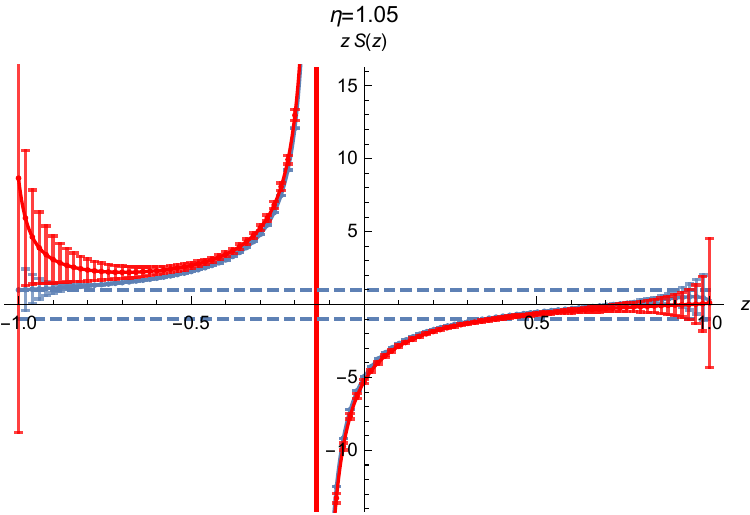}~~~
	\includegraphics[width=.31\linewidth]{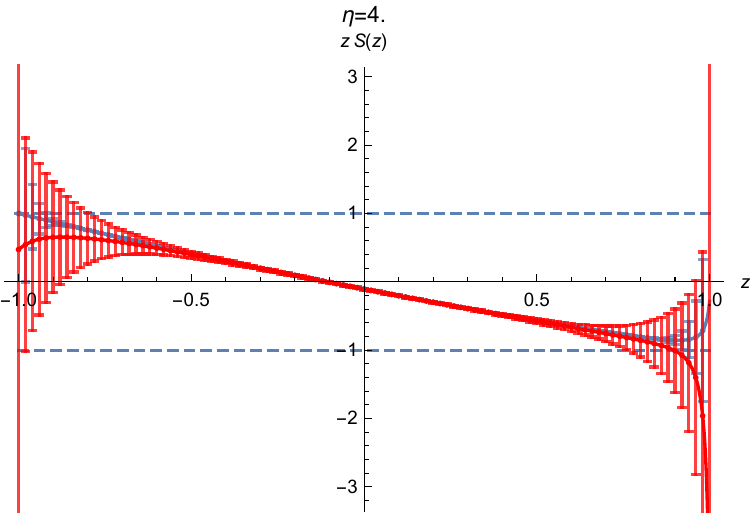}
	\caption{In red, numerical results for $zS(z)$ for real $z\in (-1, 1)$ evaluated using method I, and error bounds evaluated using  (\ref{lemmabound}), at $\eta=0.45, 1.05, 4$. In blue, results from method II as in figure \ref{fig:eta_RealLine}. 
	}
	\label{fig:cmpWmethodI}
\end{figure}

In section \ref{sec:NAC} we also introduced method I, which is more intuitive and does not rely on the presumption of number of zeros. A drawback of this method is that the explicit construction of the required $\chi(z)$ function as in  (\ref{chipnz}) is computationally expensive. In practice, to reduce error in the numerical integration, we minimize a suitably weighted combination of the absolute value of $\chi(z)$ on the unknown arc and its gradient along the unit circle. 
The resulting approximate $zS(z)$ and its error bound (computed from (\ref{lemmabound})) for real $z$ are shown in Figure \ref{fig:cmpWmethodI}, at several $\eta$ values, and compared to the results obtained using method II (as in Figure \ref{fig:eta_RealLine}). 

Note that the error bounds obtained from the two methods in Figure \ref{fig:cmpWmethodI} do not always overlap. This is caused by errors in the elastic scattering data themselves, due to finite size effects and the finite truncation level in TFFSA, as an erroneous set of scattering phases on the elastic arc may be incompatible with unitarity and analyticity constraints on $S(z)$. To reduce such errors, we would like to restrict the sampling range of the radius $R$, after which the ``known region" ${\cal I}'$ of the elastic arc is typically a set of disjoint arcs (see the domain of the purple curves in Figure \ref{fig:eta10cmppert}, for instance). This is not an issue in applying method II, but would be difficult to handle with method I, as the required $\chi$ function (small in the complement of ${\cal I}'$ on the unit circle but order 1 in the bulk of the disc) would be computationally expensive to construct.

\begin{figure}[h!]
	\centering
	\includegraphics[width=.45\linewidth]{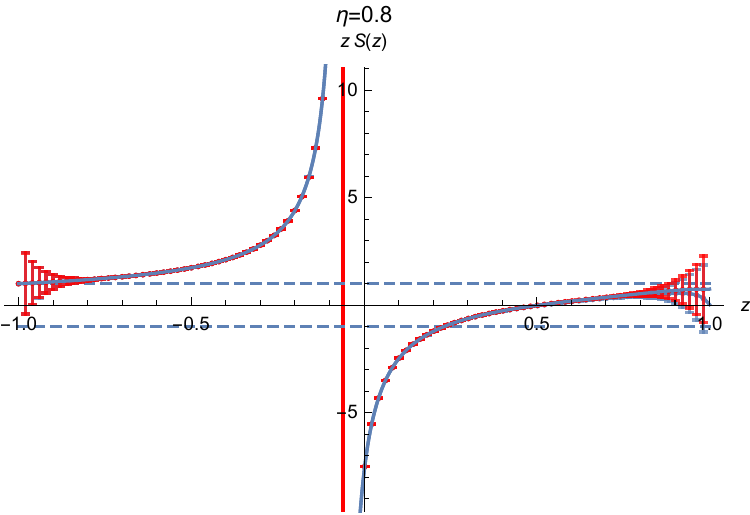}~~~~~
	\includegraphics[width=.45\linewidth]{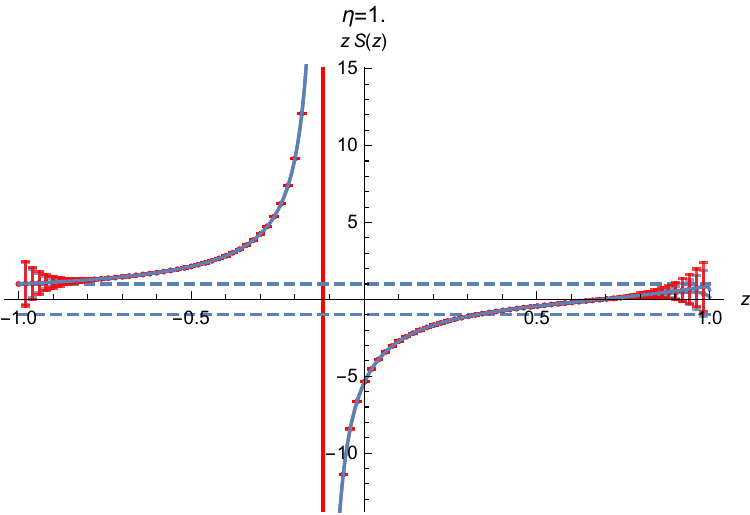}
	\centering
	\includegraphics[width=.45\linewidth]{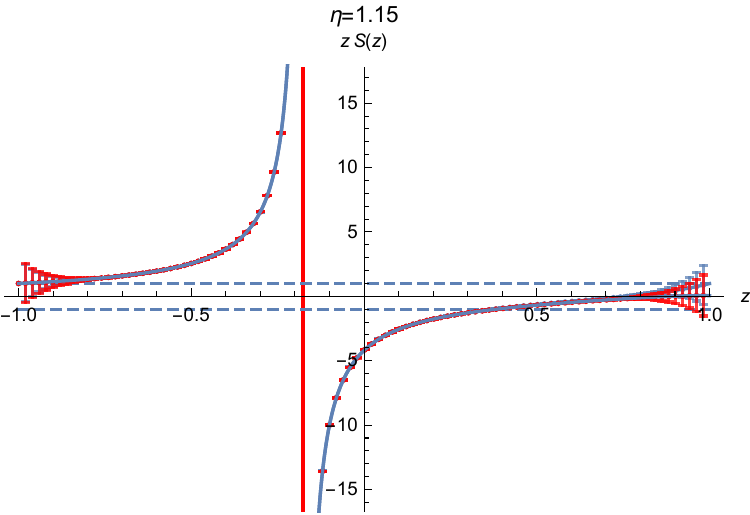}~~~~~
	\includegraphics[width=.45\linewidth]{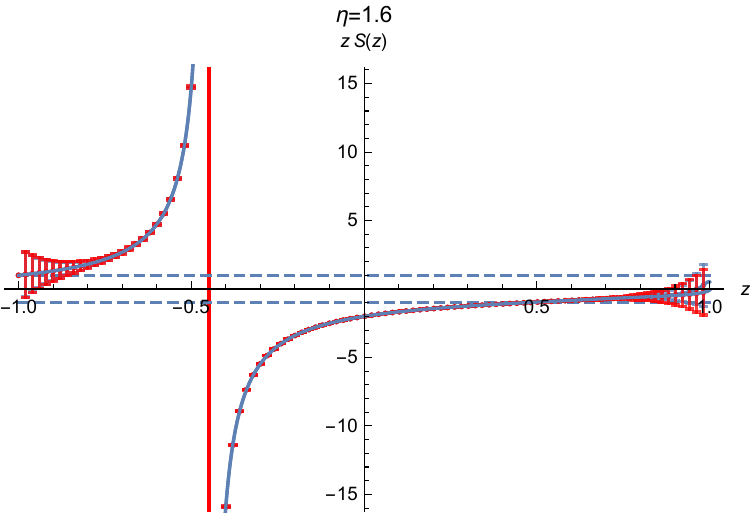}
	\caption{Error bounds on the numerical fitting of $z S(z)$ computed using method II, with the assumption of extra complex resonance $z_x'$ (in blue) and without $z_x'$ (in red).
	}
	\label{fig:cmpwwoutz_x}
\end{figure}

In applying method II, we have also considered fittings with different assumptions on the number of CDD zeros. Figure \ref{fig:cmpwwoutz_x} demonstrates the comparison of error bounds on $zS(z)$ along the real $z$-axis in the fittings with and without the extra pair of complex resonances $z_x'$ and $(z_x')^*$.

\bibliographystyle{JHEP}
\bibliography{IFTv3}

\end{document}